\documentclass[conference,compsoc]{IEEEtran}
\IEEEoverridecommandlockouts
\usepackage[table]{xcolor}
\usepackage{graphicx}
\usepackage[caption=false,font=footnotesize,labelfont=sf,textfont=sf]{subfig}

\usepackage{amsmath}
\usepackage{amssymb}
\usepackage{amsthm}
\usepackage{aliascnt}

\usepackage{booktabs}
\usepackage{multirow}

\usepackage[ruled]{algorithm2e}
\usepackage{enumitem}

\usepackage[nocompress]{cite}
\usepackage{hyperref}
\usepackage{etoolbox}
\patchcmd{\footnoterule}{0.5\columnwidth}{\columnwidth}{}{}

\newtheorem{theorem}{Theorem}

\newaliascnt{lemma}{theorem}
\newtheorem{lemma}[lemma]{Lemma}
\aliascntresetthe{lemma}

\newaliascnt{corollary}{theorem}
\newtheorem{corollary}[corollary]{Corollary}
\aliascntresetthe{corollary}

\newaliascnt{assumption}{theorem}

\aliascntresetthe{assumption}

\theoremstyle{definition}
\newaliascnt{definition}{theorem}
\newtheorem{definition}[definition]{Definition}
\aliascntresetthe{definition}

\providecommand*{\appendixautorefname}{Appendix}
\newcommand*{\appendixref}[1]{\hyperref[#1]{\mbox{\appendixautorefname~\ref*{#1}}}}

\hypersetup{
  pdftitle={AutoClip-DPZO: Practical Private Clipping Selection for Differentially Private Zeroth-Order LLM Fine-Tuning},
  pdfauthor={Xinghe Chen, Haichao Sha, Zihao Wang, Xiaokui Xiao, XiaoFeng Wang, Wei Dong},
  colorlinks,
  linkcolor={blue!70!green},
  citecolor={green!70!blue},
  urlcolor=black
}

\begin{document}
\title{\texttt{AutoClip-DPZO}: Practical Private Clipping Selection for Differentially Private Zeroth-Order LLM Fine-Tuning}

\author{
    \IEEEauthorblockN{Xinghe Chen\textsuperscript{1,*}, Haichao Sha\textsuperscript{1,*}, Zihao Wang\textsuperscript{1,*}, Xiaokui Xiao\textsuperscript{2}, XiaoFeng Wang\textsuperscript{1}, and Wei Dong\textsuperscript{1,\textdagger}}
    \IEEEauthorblockA{\textsuperscript{1}Nanyang Technological University, Singapore\qquad
                      \textsuperscript{2}National University of Singapore, Singapore}
    \thanks{\textsuperscript{*}Equal contribution.\quad
            \textsuperscript{\textdagger}Corresponding author.}
}

\maketitle

\begin{abstract}
    Differentially private zeroth-order optimization (DPZO) is a promising approach for privacy-preserving LLM fine-tuning, offering exceptional memory efficiency.
    However, its utility is highly sensitive to the clipping threshold $C$, a key DP-specific hyperparameter that controls the trade-off between noise injection and clipping bias.
    Existing DPZO methods typically tune $C$ by repeatedly evaluating candidate values and reporting the best result, which is computationally expensive and incurs additional privacy cost that must be accounted for under DP.
    Designing a practical DP procedure for selecting $C$ in DPZO is challenging because no fixed choice performs well across settings, straightforward search-based strategies suffer from poor utility and efficiency, and existing clipping-selection methods for DPSGD do not transfer well to DPZO due to its fundamentally different error regime.
    In this paper, we propose \texttt{AutoClip-DPZO}, a practical DP method that automatically and efficiently selects $C$ for DPZO.
    Our key idea is to avoid fine-tuning-based search by reducing clipping selection to a private quantile estimation problem, where the target quantile depends only on public privacy and training parameters.
    To derive this selection rule, we develop a novel characterization of DPZO error as a function of $C$ that captures both noise variance and clipping bias while remaining accurate and easy to optimize.
    Experiments show that \texttt{AutoClip-DPZO} achieves utility comparable to that obtained with the optimal $C$, while avoiding repeated fine-tuning-based search and incurring negligible additional overhead.
    Moreover, \texttt{AutoClip-DPZO} does not rely on sensitive hyperparameters.
\end{abstract}

{
\section{Introduction}

Large language model (LLM) fine-tuning is a key paradigm for adapting general-purpose LLMs to domain-specific applications, equipping them with specialized knowledge and task-solving capabilities beyond generic pretraining.
However, it also raises serious privacy concerns, as domain-specific datasets may contain private user information, proprietary records, or other sensitive content subject to regulations such as GDPR~\cite{ref1}, HIPAA~\cite{ref2}, and CCPA~\cite{ref3}.
Recent studies further show that adversaries can exploit fine-tuned LLMs through membership inference~\cite{ref4,ref5,ref6} and related data-extraction attacks~\cite{ref7,ref8} to determine whether specific examples were used in training or even recover portions of the training data.

To address these concerns, the \textit{de facto} privacy standard, \textit{differential privacy} (DP)~\cite{ref9}, has been widely incorporated into machine learning to protect sensitive data~\cite{ref10}.
Roughly speaking, DP guarantees that any single individual's contribution to the final model is statistically masked, thereby preserving privacy.
Early work applied DP to \textit{stochastic gradient descent} (SGD), yielding DPSGD~\cite{ref11}.
By clipping per-example gradients to bound sensitivity and then injecting carefully calibrated noise at each update step, DPSGD makes the training process comply with DP.
Moreover, adaptive private optimizers such as DP-AdamW~\cite{ref12} have been shown to provide more stable optimization and stronger empirical performance in LLM fine-tuning settings.
Despite their effectiveness, however, these methods can be difficult to deploy in many real-world scenarios, where regulatory, contractual, or organizational constraints may restrict or complicate the transfer of sensitive data to remote servers or third-party computing environments~\cite{ref13,ref2,ref3}.
In such settings, fine-tuning may need to be conducted in data-owner-controlled environments with limited computational resources.
GPU memory is a particularly severe bottleneck, making the aforementioned fine-tuning methods impractical for LLMs, especially when the models contain billions of parameters.
To reduce this cost, recent work has explored \textit{differentially private zeroth-order optimization} (DPZO) for LLM fine-tuning~\cite{ref14,ref15,ref16,ref17}, which estimates gradients using only $K$ random directions, where $K$ typically ranges from $16$ to $256$ and is much smaller than the original model size, and injects noise accordingly.
This approach avoids explicit backpropagation, thereby significantly reducing GPU memory overhead.
Therefore, DPZO is particularly attractive in such settings.

Despite its substantially improved memory efficiency, it has been widely observed that the performance of DPZO remains highly sensitive to the clipping threshold $C$~\cite{ref17,ref16,ref14,ref15}, a crucial DP-specific hyperparameter that bounds the contribution of each individual example and must be specified explicitly.
Roughly speaking, before noise is injected into the $K$-dimensional ZO gradients, each per-record ZO gradient is clipped using the threshold $C$, meaning that if its $\ell_2$-norm exceeds $C$, it is rescaled to have norm $C$.
This step prevents any individual record from contributing excessively to the model update: after clipping, noise scaled to $C$ is sufficient to mask the contribution of each record.
As a result, $C$ induces a trade-off between the amount of injected noise and the information loss, and thus can significantly affect model utility: when using DPZO, model performance can drop by up to $17.6$ points even when changing only $C$ by a factor of $2$.
Therefore, choosing an appropriate clipping threshold $C$ is crucial for achieving high model utility in DPZO.

However, existing DPZO methods do not address this problem in a practical way.
Instead, they repeat the entire fine-tuning procedure on the same private dataset for multiple candidate values of $C$, allocate the full privacy budget to each run, and then report the utility of the best-performing model.
This approach is problematic for two reasons.
First, it incurs substantial computational cost, since the full fine-tuning run must be repeated for each candidate $C$.
Second, DP requires accounting for the privacy loss of all runs that influence the final selection, even if some intermediate fine-tuned models are never released. Thus, $C$ selection itself incurs privacy cost: repeated search either increases the overall privacy loss or, under a fixed privacy budget, leaves less budget for training the final model.
This motivates the need for a practical method that can select $C$ efficiently while avoiding repeated privacy-consuming fine-tuning runs on the private dataset.

\noindent\textbf{Challenges.}
Selecting an appropriate $C$ for DPZO is highly challenging.
Below, we show that heuristic-based and case-specific choices of $C$, as well as methods developed for DPSGD/DP-AdamW, fail to achieve the desired utility.

\textit{First}, no fixed $C$ works well across all cases. The optimal value of $C$ can vary significantly across datasets, models, and hyperparameter settings (see \autoref{sec9}).
\textit{Second}, selecting $C$ case by case under DP is itself challenging.
One straightforward idea is to allocate a portion of the privacy budget to evaluate candidate values of $C$ and reserve the rest for the final fine-tuning run.
However, this introduces a difficult trade-off.
Reliable selection of $C$ requires a non-negligible privacy budget, as otherwise the evaluation signal becomes too noisy to reliably identify a good clipping threshold.
At the same time, spending too much budget on selection leaves insufficient privacy budget for the final DPZO fine-tuning, substantially degrading the utility of the released model.
Consequently, such a direct budget-splitting approach tends to yield poor end-to-end utility (see \autoref{sec10}).
\textit{Moreover}, another approach is to privately estimate a quantile of the $\ell_2$ norms of the ZO gradients and use it as $C$.
However, we empirically show that no fixed quantile consistently achieves strong utility.
For example, although the $0.75$ quantile works well in one setting, it causes a utility drop of up to $6.7$ points relative to the optimal choice in another.

Meanwhile, existing clipping-threshold selection techniques for DPSGD/DP-AdamW do not directly transfer to DPZO because DPZO operates in a much lower-dimensional noise regime.
In DPSGD/DP-AdamW, noise is injected into the full $d$-parameter space, where $d$ can reach billions.
Since the total noise magnitude scales with $\sqrt{d}$, the noise term typically dominates the clipping bias in practice.
In contrast, DPZO injects noise only into the low-dimensional space spanned by the $K$ random ZO gradients, where $K < 10^3 \ll d$.
As a result, the noise is substantially reduced, making clipping bias also important when selecting $C$.
This difference fundamentally changes the trade-off in clipping threshold selection.
Consequently, directly applying methods tailored to DPSGD/DP-AdamW, such as the widely adopted Automatic Clipping~\cite{ref18}, can reduce model utility by more than $10$ points.
See~\appendixref{app1} for further discussion of related work.

\subsection{Our Results}
In this work, we propose \texttt{AutoClip-DPZO}, a practical method for automatic clipping selection in DPZO fine-tuning.
Surprisingly, \texttt{AutoClip-DPZO} achieves model utility comparable to that attained using the non-DP optimal $C$, which is identified through iterative search.
In addition, \texttt{AutoClip-DPZO} incurs negligible runtime overhead.

Our high-level idea is to privately select $C$ using only a small number of sampled gradients, thereby eliminating the need for fine-tuning-based search and requiring only a single DPZO fine-tuning run.
Specifically, we reduce clipping selection to a private quantile estimation problem over the $\ell_2$ norms of the sampled gradients, where the target quantile can be determined from public privacy and training parameters.
To derive this selection rule, we analyze the DP-induced error as a function of $C$, capturing the trade-off between injected noise and clipping bias.
The main challenge lies in formulating the bias.
Through theoretical analysis and experiments, we show that the standard triangle-inequality-based upper bound on clipping bias~\cite{ref19,ref20,ref21} is too loose, leading to inaccurate selection of $C$ and reduced model utility.
In contrast, we subsequently show that the exact bias formulation is complex and coupled with the data distribution, making the resulting overall error formulation difficult to optimize.

To overcome this challenge, we develop a novel bias formulation that balances accuracy and optimizability.
We first use zeroth-order approximation and ladderization theory to transform the exact bias formulation into one that is easy to optimize.
However, this new formulation depends on a constant term specific to the setting and task.
We then eliminate this setting- and task-dependent term through theory-guided global calibration, yielding a formulation that generalizes across settings, as validated by the final end-to-end model utility.
The resulting formulation is easy to optimize and provides a more accurate characterization of the bias than standard upper bounds.
Using this formulation, \texttt{AutoClip-DPZO} can privately determine an appropriate clipping threshold without repeatedly training candidate models, preserving DPZO's computational advantages while avoiding unnecessary privacy cost.

In summary, our contributions are threefold:

\noindent
\vspace{2pt}\noindent$\bullet$\textbf{~Practical Clipping Selection:}
We identify clipping selection as a practical bottleneck in DPZO: existing fine-tuning-based search incurs substantial computation and extra privacy loss.
We develop \texttt{AutoClip-DPZO}, which privately selects an appropriate $C$ without repeatedly fine-tuning candidate models.

\vspace{2pt}\noindent$\bullet$\textbf{~Theoretical Analysis:}
To enable this efficient selection, we provide a novel theoretical formulation of the DP error in DPZO as a function of $C$.
This formulation accurately captures the DP error while remaining easy to optimize.
Using this formulation, we reduce $C$ selection to quantile estimation, where the target quantile depends only on public parameters, such as the privacy parameters $(\varepsilon,\delta)$, the number of directions $K$, and the batch size $b$.

\vspace{2pt}\noindent$\bullet$\textbf{~Experimental Results:}
We conduct comprehensive experiments to evaluate \texttt{AutoClip-DPZO}.
In our utility experiments, our method achieves performance comparable to non-DP search for the optimal $C$, with an average utility drop of only $0.48$ points, and even outperforms it in some cases.
Compared with DP baselines further strengthened by non-private tuning of their respective hyperparameters, \texttt{AutoClip-DPZO} outperforms the best-performing baseline in each setting by up to $1.8$ points, with an average error reduction of $3.7\%$.
In terms of efficiency, our method incurs only $3\%$--$27\%$ additional runtime overhead compared with fine-tuning using a predefined $C$.
Compared with search-based methods, it achieves up to a $6.90\times$ speedup.

\section{Background}
\label{sec1}

\subsection{LLM Fine-Tuning}
\label{sec2}

LLMs are typically pretrained on broad corpora to acquire general language understanding and generation capabilities.
To adapt them to downstream applications, practitioners often fine-tune them on task- or domain-specific datasets.
Fine-tuning has become standard for applications such as medical question answering, legal and financial analysis, enterprise assistants, code generation, and personalized recommendations~\cite{ref22}.
Training on target-domain data can improve domain knowledge, task-following behavior, and alignment with application-specific output formats.

However, the domain-specific data that makes fine-tuning useful also introduces privacy risks.
In practice, fine-tuning datasets may contain sensitive user information, proprietary business records, confidential documents, or regulated data.
Once such data is used for training, a released fine-tuned model may unintentionally memorize parts of the training set.
Prior work has shown that adversaries can exploit this memorization through membership inference~\cite{ref4,ref5,ref6} or data extraction attacks~\cite{ref7,ref8}, potentially revealing whether a record was used in training or exposing sensitive training content.
These risks are particularly concerning in high-stakes domains or when fine-tuning involves third-party platforms.

Therefore, privacy-preserving LLM fine-tuning aims to retain the benefits of private-data adaptation while limiting what the model reveals about individual training examples.

\textit{Notation.}
We denote the private fine-tuning dataset by $D = \{x_1, \ldots, x_n\} \in \mathcal{D}$, where $D$ contains $n$ samples and $\mathcal{D}$ denotes the universe of all datasets.
Let $B \subseteq D$ denote a batch of size $b$ sampled from $D$, such that each sample has marginal inclusion probability $b/n$.
Let $\theta$ denote the model parameters to be fine-tuned.
We write the model loss as $L(\theta,\cdot)$, and denote the corresponding gradients with respect to $\theta$ on a batch and a sample by $\nabla_{\theta} L(\theta,B)$ and $\nabla_{\theta} L(\theta,x)$, respectively.
We use $\|\cdot\|_2$ to denote the $\ell_2$ norm of a vector.
When $t$ appears as a subscript, $(\cdot)_t$ denotes the value of a variable at the $t$-th iteration.
We use $\widehat{(\cdot)}$, $\overline{(\cdot)}$, and $\widetilde{(\cdot)}$ to denote an estimator, a clipped version, and a privatized quantity, respectively.
We denote by $[L, U]$ the set of integers from $L$ to $U$.

{
\subsection{Differential Privacy}
\label{sec3}

Differential privacy formalizes the intuition that the participation of any single record should have only a limited influence on the output of an algorithm.
The standard $(\varepsilon,\delta)$-DP model is defined as follows:

\begin{definition}[Standard DP~\cite{ref9}]
     For $\varepsilon,\delta > 0$, a mechanism $\mathcal{M}$ is $(\varepsilon,\delta)$-DP if, for any pair of neighboring datasets $D \sim D'$ that differ by one record and any measurable output event $S$, the following holds:
    \begin{equation*}
        \Pr[\mathcal{M}(D) \in S] \leq e^\varepsilon \cdot \Pr[\mathcal{M}(D') \in S] +\delta.
    \end{equation*}
\end{definition}
Roughly speaking, a smaller $\varepsilon$ provides stronger privacy, while $\delta \ll 1/n$ bounds the privacy failure probability.

A commonly used DP mechanism is the \textit{Gaussian mechanism}, which injects Gaussian noise according to the sensitivity of the function, defined as follows:
\begin{lemma}[Gaussian Mechanism~\cite{ref9}]
\label{lem1}
Let $f : \mathcal{D} \to \mathbb{R}^d$ be a function with $\ell_2$-sensitivity $\Delta_2(f)=$ $\sup_{D \sim D'}$ $\| f(D)-f(D') \|_2$, and consider the Gaussian mechanism
$
    \mathcal{M}(D)=f(D)+\mathcal{N}(0,\sigma^2 I_d).
$
Then, for any $\varepsilon \in (0,1)$ and $\delta \in (0,1)$, $\mathcal{M}$ satisfies $(\varepsilon,\delta)$-DP whenever
$
    \sigma \ge \Delta_2(f)\sqrt{2\log(1.25/\delta)}/{\varepsilon}.
$
\end{lemma}

In addition to noise injection, DP can also be achieved via randomized response.
It has been shown that the quantile selection problem, which aims to output the value at a specified quantile of a set of values, can be solved under $(\varepsilon,0)$-DP using the \textit{exponential mechanism}~\cite{ref23}, with strong utility guarantees~\cite{ref24,ref25,ref26}.
For details on the exponential mechanism for quantile selection, see \appendixref{app9}.

DP mechanisms naturally enjoy composition properties.
Roughly speaking, \textit{sequential composition}~\cite{ref9} makes it possible to compose mechanisms, with privacy loss growing linearly in the number of composed mechanisms.
Moreover, for iterative mechanisms that use only a subsampled subset of the original dataset $D$, as is typically the case in model training, the \textit{Moments Accountant}~\cite{ref11,ref27} provides a much tighter composition analysis than sequential composition.
For more DP background, see \appendixref{app8}.
\subsection{Privacy-Preserving LLM Fine-Tuning}
\label{sec4}

We next review DP in LLM fine-tuning, where each training update is privatized by clipping per-sample contributions and adding noise.
Existing approaches broadly fall into first-order methods, which rely on backpropagation, and zeroth-order methods, which use forward-only loss evaluations to reduce memory overhead.

\subsubsection{First-Order DP LLM Fine-Tuning and Its Limitations}
\label{sec5}

Non-DP LLM fine-tuning adapts a pretrained model to a downstream task through gradient-based updates, such as $\theta_{t+1} = \theta_t - \eta \nabla_{\theta} L(\theta_t, B_t)$, where $\eta$ is the learning rate and $B_t$ is a batch randomly drawn at step $t$.
When the fine-tuning data contain sensitive information, DPSGD~\cite{ref11} protects each sample by \textit{clipping} its gradient and adding Gaussian noise.
Adaptive optimizers such as DP-AdamW~\cite{ref12} build on the same principle while improving optimization stability and empirical performance in LLM fine-tuning.
For each sample $x \in B_t$, DPSGD computes $\nabla_{\theta_t} L(\theta_t, x)$ and clips it to the $\ell_2$ bound $C > 0$:
\begin{equation*}
    \overline{\nabla_{\theta_t}} L(\theta_t, x)
    =
    \nabla_{\theta_t} L(\theta_t, x)\Big/\max\left\{1, \frac{\|\nabla_{\theta_t} L(\theta_t, x)\|_2}{C}\right\}.
\end{equation*}
It then aggregates the clipped gradients and adds Gaussian noise:
\begin{equation*}
\begin{aligned}
    \widetilde{\nabla_{\theta_t}}
    &=
    \frac{1}{b}\left(
        \sum_{x \in B_t} \overline{\nabla_{\theta_t}} L(\theta_t, x)
        +
        \mathcal{N}\!\left(0, \sigma^2 C^2 I_d\right)
    \right),\\
    \theta_{t+1}
    &=
    \theta_t - \eta \widetilde{\nabla_{\theta_t}},
\end{aligned}
\end{equation*}
where $\sigma$ is the noise multiplier and $I_d$ is the identity matrix.
Here, clipping bounds the sensitivity of the aggregated gradient, allowing the Gaussian mechanism to achieve DP by adding noise scaled to $C$.
Then, for the iterative access to the private dataset during model fine-tuning, the Moments Accountant is used to compose the overall privacy loss.

Although first-order DP methods provide rigorous $(\varepsilon,\delta)$-DP guarantees and can achieve strong utility, they remain expensive for LLM fine-tuning.
The main bottleneck is that DP requires per-sample gradient computation before aggregation, which is much more memory-intensive than non-DP training.
For LLMs with millions or billions of parameters, this overhead can make first-order DP fine-tuning difficult to deploy, especially when private data cannot be transferred to large remote computing platforms and must instead be processed locally.
This motivates zeroth-order (ZO) approaches, which replace backpropagation with forward-only loss evaluations and substantially reduce memory overhead.

\subsubsection{Non-DP ZO Optimization for LLM Fine-Tuning}
\label{sec6}

ZO optimization addresses this limitation by replacing exact gradients with approximations derived from forward-only loss evaluations, a strategy shown to be effective for LLM fine-tuning~\cite{ref28}.
Concretely, for a random direction $z \sim \mathcal{N}(0,I_d)$ and a small radius $\mu > 0$, it uses the classical two-point estimator to compute the scalar loss difference.
\begin{equation*}
    \Delta_{z}(\theta, x)
    =
    \frac{L(\theta + \mu z, x) - L(\theta - \mu z, x)}{2\mu},
\end{equation*}
which yields the ZO gradient estimate
\begin{equation*}
    \widehat{\nabla}_{z} L(\theta, x)
    =
    \Delta_{z}(\theta, x)\cdot z,\
    \widehat{\nabla}_{z} L(\theta, B)
    =
    \frac{1}{b}\sum_{x \in B}\widehat{\nabla}_{z} L(\theta, x).
\end{equation*}
It can be shown that the ZO estimator is unbiased with respect to the random direction and is thus effective~\cite{ref28}.

\subsubsection{DPZO and Aggregation over \texorpdfstring{$K$}{K} Random Directions}
\label{sec7}

ZO gradients have a convenient structure for DP: the random direction $z$ is drawn from public randomness, while the data-dependent quantity is the scalar $\Delta_z(\theta, x)$, or $\Delta_z(\theta, B)$ for a batch.

The original DPZO method~\cite{ref16} uses a single direction and clips the per-sample scalar loss difference.
The clipped values are averaged over a batch, perturbed with one-dimensional Gaussian noise, and used to update the model along that direction, thereby preserving privacy by \autoref{lem1}.
However, using only one direction can lead to high variance.
To reduce this variance, multi-direction DPZO~\cite{ref14,ref17} samples $K$ i.i.d.\ directions $\{z_k\}_{k=1}^K$ and computes, for each sample $x$,
\begin{equation*}
    \Delta_{K}(\theta, x)
    =
    \frac{1}{K}\big(\Delta_{z_1}(\theta, x), \ldots, \Delta_{z_K}(\theta, x)\big)
    \in \mathbb{R}^K.
\end{equation*}
It then clips this $K$-dimensional vector with $C > 0$\footnote{In~\cite{ref14}, clipping is performed separately for each dimension.
Later work~\cite{ref17} shows that clipping the aggregated $K$-dimensional ZO gradient strictly reduces clipping error. Here, we follow the formulation in~\cite{ref17} throughout this study.}:
\begin{equation*}
    \overline{\Delta}_{K}(\theta, x)
    =
    \Delta_{K}(\theta, x)\Big/\max\left\{1, \frac{\|\Delta_{K}(\theta, x)\|_2}{C}\right\}.
\end{equation*}
Next, it adds Gaussian noise in the $K$-dimensional space:
\begin{equation*}
    \widetilde{\Delta}_{K}(\theta, B)
    =
    \frac{1}{b}\left(
        \sum_{x \in B}\overline{\Delta}_{K}(\theta, x)
        +
        \mathcal{N}\!\left(0, \sigma^2 C^2 I_K\right)
    \right).
\end{equation*}
The corresponding parameter update is
\begin{equation*}
    \widetilde{\nabla_{\theta_t, K}}
    =
    \sum_{k=1}^{K} \big[\widetilde{\Delta}_{K}(\theta_t, B)\big]_k \, z_k,
    \qquad
    \theta_{t+1} = \theta_t - \eta \widetilde{\nabla_{\theta_t, K}}.
\end{equation*}
Then, the Moments Accountant can be used to compose the privacy loss over iterative access to the private dataset.
The case $K = 1$ recovers the original DPZO with a single-direction update.
Here, a moderate $K > 1$ reduces the variance caused by the randomness of the directions while requiring DP noise only in a $K$-dimensional space rather than across all model parameters.
As a result, multi-direction DPZO can match or even outperform the state-of-the-art (SOTA) first-order DP-AdamW~\cite{ref12} in some settings while retaining the memory benefits of ZO methods~\cite{ref17}.

Notably, recent work shows that the intrinsic randomness of ZO directions provides no inherent privacy advantage~\cite{ref29}, except under restrictive assumptions and extreme parameter regimes~\cite{ref30}.
Thus, practical DPZO methods still rely on standard gradient clipping and noise injection, motivating our evaluation of this approach.
}

\section{Motivation and Challenges}
\label{sec8}

{
\subsection{Motivation for Private Clipping Selection}
\label{sec9}

As discussed in \autoref{sec4}, DPZO reduces the memory overhead of private LLM fine-tuning by injecting noise only in the low-dimensional space of $K$ ZO gradients.
However, this different noise regime also makes the clipping threshold $C$ much more critical.

Unlike prior first-order methods such as DPSGD, where $C$ is often effective within a relatively small range (e.g., $[2^{-3}, 2^{0}]$)~\cite{ref11,ref18}, empirical studies on DPZO suggest that the best-performing threshold can lie in a much wider range, roughly $[2^{-2}, 2^{8}]$~\cite{ref14,ref17}.
Consistent with this observation, we find that the optimal $C$ can differ by up to $17.1\times$ across datasets, even when the model and other hyperparameters are fixed.
Moreover, even a $2\times$ change in $C$ can reduce model utility by up to $17.6$ points.
Thus, a globally fixed threshold cannot reliably support DPZO fine-tuning.
A natural alternative is to choose $C$ case by case using a predefined quantile of the $\ell_2$ norms of the ZO gradients.
However, no fixed quantile consistently performs well either, leading to utility drops of up to $6.7$ points in our experiments.
More empirical details on the failures of fixed $C$ and fixed-quantile choices are provided in \autoref{tab3}.

To understand why DPZO is particularly sensitive to $C$, we examine the DP-induced error.
Let $\Delta_{K} := \Delta_{K}(\theta, B)$ denote the unclipped batch mean in the $K$-dimensional ZO-gradient space, let $\overline{\Delta}_{K}$ denote the corresponding clipped batch mean, and let $\widetilde{\Delta}_{K}$ denote the privatized estimate after Gaussian perturbation.
By the triangle inequality,
\begin{equation}
\label{eq1}
    \left\| \widetilde{\Delta}_{K} - \Delta_{K} \right\|_2
    \le
    \underbrace{\left\| \widetilde{\Delta}_{K} - \overline{\Delta}_{K} \right\|_2}_{\text{Noise error}}
    +
    \underbrace{\left\| \overline{\Delta}_{K} - \Delta_{K} \right\|_2}_{\text{Clipping bias}}.
\end{equation}
The first term comes from the Gaussian perturbation.
Its expected magnitude increases with both $C$ and the square root of the noise dimension:
\begin{equation}
\label{eq2}
    \mathbb{E}\!\left[\text{Noise}(C)\right]
    =
    \mathbb{E}\!\left[
        \left\|
            \mathcal{N}\!\left(0, \frac{\sigma^2 C^2 I_{K}}{b^2}\right)
        \right\|_2
    \right]
    \le \frac{\sigma C \sqrt{K}}{b},
\end{equation}
where the bound follows from Jensen's inequality.

The second term is the clipping bias.
It depends on how much the per-example ZO gradients are rescaled by clipping:
\begin{equation*}
    \begin{aligned}
        \text{Bias}(C)
        =
        \frac{1}{b}
        \Biggl\|
            \sum_{x \in B}
            \Biggl(
                \frac{\Delta_{K}(\theta, x)}
                {\max\!\left\{
                    1,
                    \frac{\left\| \Delta_{K}(\theta, x) \right\|_2}{C}
                \right\}}
                - \Delta_{K}(\theta, x)
            \Biggr)
        \Biggr\|_2
        .
    \end{aligned}
\end{equation*}
Although this expression is data-dependent and does not immediately yield a closed-form selection rule, it reveals the key trade-off.
Increasing $C$ reduces clipping and therefore decreases the bias, but it also increases the Gaussian noise calibrated to $C$.
Thus, $C$ directly controls the noise--bias trade-off that determines DPZO utility.
This trade-off also exists in first-order DP methods, but DPZO operates in a fundamentally different regime.
In DPSGD and DP-AdamW~\cite{ref11,ref12}, Gaussian noise is injected into the full $d$-dimensional parameter space, where $d$ can reach millions or billions for LLMs.
In contrast, DPZO injects noise only into the $K$-dimensional ZO-gradient space, where typically $K \in [10^{1},10^{3}] \ll d$.
Consequently, the noise term is much larger in first-order DP training and often dominates the effect of clipping bias.
In DPZO, however, the low-dimensional noise makes the bias term much more important.
This explains why DPZO prefers a larger $C$ and why its utility is highly sensitive to the choice of $C$.

In existing DPZO workflows~\cite{ref14,ref15,ref16,ref17}, $C$ is usually treated as a hyperparameter and selected by repeating fine-tuning with multiple candidate thresholds.
This approach is unsatisfactory in terms of both efficiency and privacy.
First, DPZO's wide candidate range for $C$ requires many fine-tuning runs.
This cost is amplified in multi-direction DPZO, which already trades runtime for memory by computing $K$ ZO gradients per step.
Additional threshold search therefore further reduces the practicality of DPZO for resource-constrained data owners.
Second, because each candidate evaluation depends on the same private dataset, the privacy loss of all such evaluations must be accounted for, even if only the final selected model is released.
This leaves less privacy budget for the final fine-tuning run and can degrade the utility of the released model.
These observations motivate a case-dependent and fully private strategy for selecting $C$ without repeatedly fine-tuning on the private dataset.
Next, we show that a direct privacy-preserving search over candidate thresholds faces an inherent privacy--utility trade-off.
\subsection{Challenges Revealed by a Strawman Search-and-Release Approach}
\label{sec10}
The previous section shows that DPZO requires a case-dependent choice of the clipping threshold $C$.
A natural idea is therefore to search over candidate thresholds under DP and release the model trained with the best one.
We show that such fine-tuning-based search is problematic because it consumes both computation and privacy budget.

Suppose we are given a candidate set $\{C_m\}_{m=1}^M$.
The most direct DP-compliant approach is to run a separate DP fine-tuning session for each candidate threshold and select the best-performing model.
Under a fixed target privacy budget $(\varepsilon,\delta)$, this requires accounting for $M$ fine-tuning runs, effectively increasing the total number of accounted training iterations by a factor of $M$.
Under the Moments Accountant, this substantially increases the required noise level.
For example, even with a modest $M = 10$ candidates covering the range $[2^{-2}, 2^{8}]$, the total number of private training iterations increases tenfold.
As a result, the required noise multiplier $\sigma$ increases by approximately $\sqrt{10}$.
Since $\sigma$ directly controls the Gaussian perturbation in~\eqref{eq2}, this additional privacy composition can significantly degrade model utility.
As shown in \autoref{tab1}, increasing the noise multiplier from $3.10$ to $9.39$ reduces model utility by up to $10.32$ points.

\begin{table}[ht]
\centering
\caption{Test accuracy across privacy budgets.}
\resizebox{0.95\linewidth}{!}{
\begin{tabular}{c|c|cc}
\toprule
\multirow{1}{*}{\textbf{Model}} &\multirow{1}{*}{\textbf{Noise Multiplier}}
& \multicolumn{1}{c}{\textbf{SST-2}}  & \multicolumn{1}{c}{\textbf{SST-5}}  \\
\midrule
\multirow{2}{*}{\textbf{RoBERTa-Large}}
&$\sigma=3.10$   &93.58 &49.70\\
&$\sigma=9.39$   &91.06 &40.00\\
\midrule
\multirow{2}{*}{\textbf{OPT-350M}}
&$\sigma=3.10$   &89.33 &68.67\\
&$\sigma=9.39$   &83.60  &58.35\\
\bottomrule
\end{tabular}
}
\label{tab1}
\end{table}

The key drawback of this direct search-and-release approach is that most of the privacy budget is spent on candidate models that are ultimately discarded.
To reduce this waste, we consider a refined two-stage strawman protocol that separates threshold search from final model release.
Specifically, the protocol introduces a budget-allocation factor $\alpha \in (0,1)$.
In the first stage, it uses the search budget $(\alpha \varepsilon, \alpha \delta)$ to evaluate all candidate values of $C$ and select a promising one.
In the second stage, it uses the remaining budget $((1-\alpha)\varepsilon, (1-\alpha)\delta)$ to fine-tune and release the final model with the selected threshold.
Within the first stage, the Moments Accountant can be used to compose the privacy loss across candidates.
Then, by sequential composition across the two stages, the full protocol satisfies $(\varepsilon,\delta)$-DP.
The formal algorithm is shown in \autoref{alg2}.

\noindent\textbf{Limitations of the strawman approach.}
Although the two-stage strategy avoids directly releasing one of the search-stage models, it still faces two major limitations.

(i) High computational overhead.
Even a coarse logarithmic grid over the DPZO threshold range requires evaluating multiple candidate values of $C$.
Since each candidate evaluation involves a separate DP fine-tuning run, the search overhead grows linearly with $M$.
This overhead is particularly costly for multi-direction DPZO.
As a result, fine-tuning-based threshold search can substantially reduce the practicality of DPZO for resource-constrained data owners.

(ii) Sensitive to hyperparameters.
The two-stage strategy introduces a new hyperparameter, the budget-allocation factor $\alpha$.
If $\alpha$ is too large, too little privacy budget remains for the final fine-tuning stage, directly harming the utility of the released model.
If $\alpha$ is too small, the search stage becomes too noisy to reliably identify a good threshold.
Moreover, tuning $\alpha$ based on private-data performance incurs additional privacy costs under DP, further reducing the available privacy budget.

These limitations show that case-dependent clipping-threshold selection in DPZO cannot be handled satisfactorily by straightforward fine-tuning-based DP search.
This motivates a theory-guided DP mechanism that selects an appropriate $C$ efficiently while preserving the privacy budget for the final model.
}

\section{Technical Roadmap}
\label{sec11}

Motivated by the above discussion in \autoref{sec8} on the importance of and challenges in clipping-threshold selection, we aim to design a DP mechanism for selecting an appropriate $C$.
Below, we provide an intuitive overview of our technical design.
Some notation and statements are kept intentionally informal for exposition.
A more rigorous and detailed analysis is provided in \autoref{sec12} and \appendixref{app3}.

To select a good clipping threshold under DP, the first task is to characterize the total error as a function of $C$.
An ideal error formulation should satisfy two properties: (1) it should accurately capture the total error arising from noise variance and clipping bias, and (2) it should be easy to optimize so that a clipping threshold can be practically derived under DP.
As noted in \autoref{sec9}, the error has two components: clipping bias and Gaussian noise.
Compared with the noise, the bias is more challenging to characterize because it depends on an unknown underlying distribution.

\noindent\textbf{Stage 1: There is a trade-off between accuracy and optimizability.}
To characterize the bias, we begin by adopting a widely used triangle-inequality-based upper bound on the bias of clipping mechanisms from the DP literature~\cite{ref19,ref20,ref21}.
Roughly speaking, this bound considers the worst-case scenario in which all clipping residuals are perfectly aligned and therefore directly sums their $\ell_2$ norms, yielding $\mathcal{F}(C)$, an upper bound on the bias.
Combining $\mathcal{F}(C)$ with the noise term yields an error formulation $\mathcal{E}_{\mathcal{F}}(C)$, which can be optimized by differentiating with respect to $C$, thereby reducing the problem to quantile selection under DP.
Since quantile estimation has been widely studied in the DP literature~\cite{ref24,ref25,ref26}, the error formulation $\mathcal{E}_{\mathcal{F}}(C)$ satisfies property~(2).
However, as shown empirically in \autoref{sec20}, this formulation is not sufficiently accurate to satisfy property~(1) because it relies on a strict perfect-alignment upper bound on the bias, an assumption that is often too strong for real fine-tuning datasets.
To better satisfy property~(1), we provide a more accurate formulation of the bias, denoted by $\mathcal{G}(C)$.
However, since $\mathcal{G}(C)$ is tightly coupled to the data-dependent exact gradients, combining it with the noise term would yield an error objective, $\mathcal{E}_{\mathcal{G}}(C)$, that is difficult to optimize.
As a result, directly using $\mathcal{G}(C)$ in the total error formulation inevitably fails to satisfy property~(2).

Thus, $\mathcal{F}(C)$ is easy to optimize but overly conservative, whereas $\mathcal{G}(C)$ is more accurate but difficult to optimize.

\noindent\textbf{Stage 2: Formulating an error objective to balance accuracy and optimizability.}
Next, we aim to formulate a bias estimate $\mathcal{R}(C)$, yielding the error objective $\mathcal{E}_{\mathcal{R}}(C)$ that balances accuracy and optimizability.
\textbf{Our goals.}
More precisely, the goal is to construct $\mathcal{R}(C)$ such that it provides a tighter estimate than $\mathcal{F}(C)$ while remaining an upper bound on the actual clipping bias at the minimizer of the corresponding error formulation.
At the same time, the resulting error objective $\mathcal{E}_{\mathcal{R}}(C)$ must be easy to optimize.
Intuitively, requiring such a moderate bias estimate would yield $C_{\mathcal{G}}^* \leq C_{\mathcal{R}}^* \leq C_{\mathcal{F}}^*$, where these are the optimal thresholds derived from the three error objectives, respectively.

\emph{Step 1: Decompose the exact objective.} We approach this problem by first rewriting $\mathcal{G}(C)$ in terms of $\mathcal{F}(C)$, which shows that the unknown term in $\mathcal{E}_{\mathcal{G}}(C)$ is precisely $\rho(C) := \mathcal{G}(C) / \mathcal{F}(C)$.
Here, $\rho(C)$ is an alignment factor that reflects how much looser $\mathcal{F}(C)$ is than $\mathcal{G}(C)$.
Since $\rho(C)$ is tightly coupled with the underlying data distribution, the theory only guarantees that $\rho(C) \in [0,1]$.
\emph{Step 2: Zeroth-order approximation.} Therefore, a natural idea is to use a zeroth-order approximation of $\rho(C)$ so that it becomes a constant term $r \in (0,1]$.
This approximation naturally yields $\mathcal{R}(C) := r \cdot \mathcal{F}(C)$, which always satisfies $C_{\mathcal{R}}^* \leq C_{\mathcal{F}}^*$ because $\mathcal{R}(C)$ is tighter than $\mathcal{F}(C)$ for any $r \in [0,1]$.
Meanwhile, the resulting error objective $\mathcal{E}_{\mathcal{R}}(C)$ yields a closed-form expression for $C_{\mathcal{R}}^*$, which depends on the value of $r$.
However, a key issue is that if we use an arbitrarily small calibration of $r$, then $C_{\mathcal{G}}^* \leq C_{\mathcal{R}}^*$ is not guaranteed.
Thus, $r$ requires careful calibration.
\emph{Step 3: Calibrate the unknown constant term.} To address this issue, we discretize $\rho(C)$ into a finite ladder to construct a candidate set of $r$ values.
We then prove that, under mild monotonicity conditions, a \textit{self-consistent} ladder calibration of $r$ guarantees that $C_{\mathcal{G}}^* \leq C_{\mathcal{R}}^*$.
Once $r$ is calibrated, $C_{\mathcal{R}}^*$ can be obtained under DP.
At this point, we have nearly achieved all the goals, but the above calibration is still setting- and task-dependent, which is undesirable.
\emph{Step 4: Eliminate setting dependence.} To close this final gap, we first replace the calibration of $r$ with carefully formulated global parameters across settings, thereby eliminating setting dependence.
\emph{Step 5: Eliminate task dependence.} Second, we eliminate task dependence by empirically showing that these global parameters are highly concentrated across tasks.
This completes the final steps toward an error objective $\mathcal{E}_{\mathcal{R}}(C)$ that balances accuracy and optimizability.

\section{Toward Practical Clipping Selection}
\label{sec12}

{
\subsection{Tension Between Accuracy and Optimizability}
\label{sec13}

The earlier discussion shows that fine-tuning-based search for $C$ is costly and privacy-inefficient (see \autoref{sec10}).
This motivates a theory-guided selection rule that estimates a good threshold directly from ZO-gradient statistics, without repeatedly fine-tuning candidate models.
However, such a rule requires an error objective that is both sufficiently accurate and tractable to optimize under DP.
In this section, we show that these two requirements are in tension.

Let $g_x := \Delta_K(\theta, x)$ denote the per-sample $K$-dimensional ZO gradient.
A natural first step is to upper-bound this term using the triangle inequality:
\begin{equation}
\label{eq3}
    \begin{aligned}
        \mathrm{Bias}(C)
        &= \frac{1}{b} \left\| \sum_{x \in B} \left( \overline{g}_{x}(C) - g_x \right) \right\|_2 \\
        &\le \frac{1}{b} \sum_{x \in B} \left\| \overline{g}_{x}(C) - g_x \right\|_2 \\
        &= \frac{1}{b} \sum_{x \in B} \max\!\left(0, \left\| g_x \right\|_2 - C\right):=\mathcal{F}(C).
    \end{aligned}
\end{equation}
This formulation is conservative: the inequality is tight only in the worst case where all clipping residuals $\{\overline{g}_{x}(C) - g_x\}_{x \in B}$ are positively aligned.
Combining this upper bound with the Gaussian noise term in~\eqref{eq2} gives the following error objective:
\begin{equation}
    \label{eq4}
    \mathcal{E}_{\mathcal{F}}(C) := \frac{1}{b} \left( \sum_{x \in B} \max\!\left(0, \left\| g_x \right\|_2 - C\right) + \sigma C \sqrt{K} \right).
\end{equation}
The advantage of $\mathcal{E}_{\mathcal{F}}(C)$ lies in its tractability.
Differentiating it with respect to $C$ yields the following closed-form rule.

\begin{lemma}[Optimal Threshold via the Bias Upper Bound]
\label{lem2}
Consider a batch of size $b$ and a $K$-dimensional DPZO estimator.
The error objective $\mathcal{E}_{\mathcal{F}}(C)$ is minimized at
\begin{equation*}
    C_{\mathcal{F}}^* = \operatorname{Quantile}\left( \{\|g_x\|_2\}_{x \in B}, \, [1-\lambda]_{[0,1]} \right),
\end{equation*}
where $\left[\cdot\right]_{[0,1]} := \min\{1,\max\{0,\cdot\}\}$ and $\lambda := \sigma \sqrt{K} / b$.
\end{lemma}

All proofs are in \appendixref{app7}.

\begin{figure}[ht]
    \centering
    \includegraphics[width=\columnwidth]{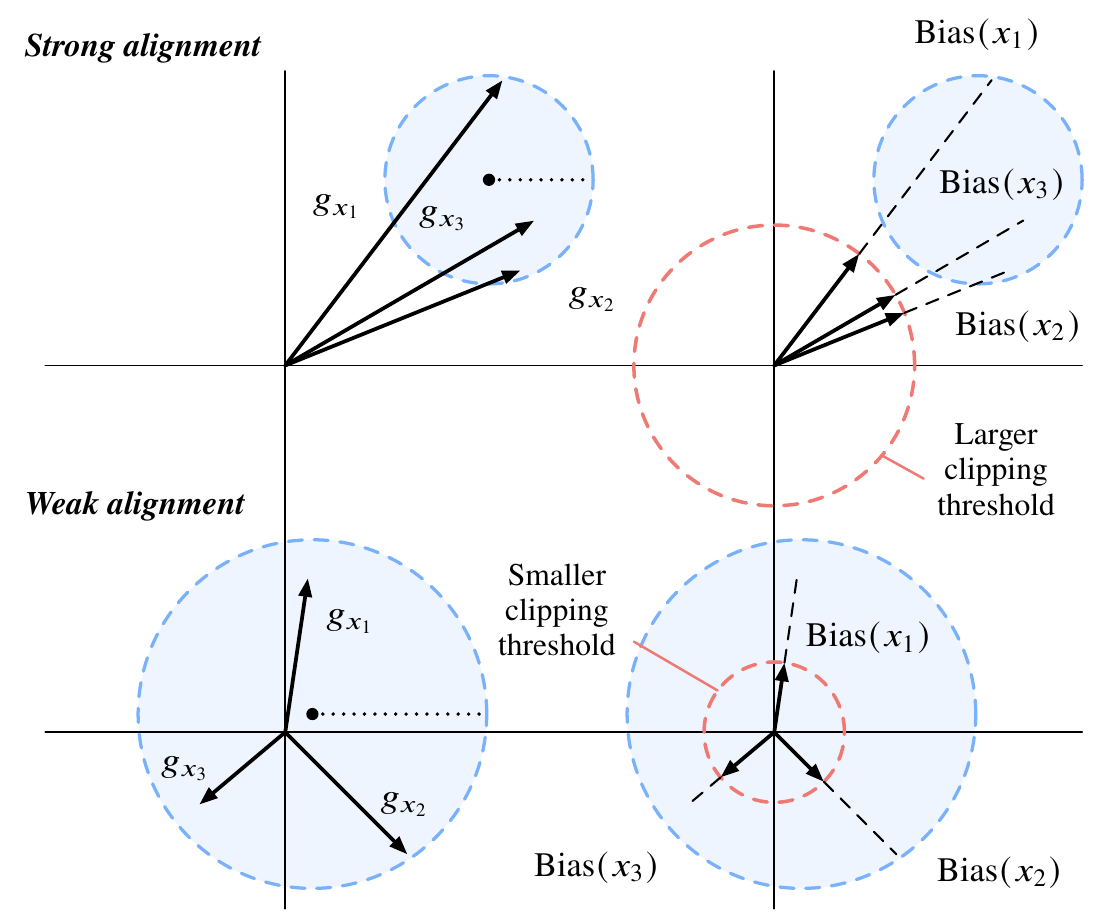}
    \caption{Illustration of the difference in clipping-bias residuals under strong and weak alignment.}
    \label{fig1}
\end{figure}

However, this optimizability comes at the cost of accuracy.
Empirically, the threshold induced by \autoref{lem2} is suboptimal in many settings, as shown in \autoref{sec20}.
The reason is that $\mathcal{F}(C)$ ignores the spatial alignment of clipping residuals.
As illustrated in \autoref{fig1}, when residuals are weakly aligned, they can partially cancel each other after aggregation.
In this case, the true clipping bias can be much smaller than the triangle-inequality upper bound.
As a result, $\mathcal{E}_{\mathcal{F}}(C)$ may overestimate the bias and favor an overly large $C_{\mathcal{F}}^*$.

To improve accuracy, one could instead preserve the exact alignment information by using the clipping-bias expression directly:
\begin{equation}
    \label{eq5}
    \mathcal{G}(C) := \frac{1}{b}\left\| \sum_{x \in B} \left( \overline{g}_x(C) - g_x \right) \right\|_2, \
    \mathcal{E}_{\mathcal{G}}(C) := \mathcal{G}(C) + \lambda C,
\end{equation}
where $\lambda := \sigma\sqrt{K}/b$.
Compared with $\mathcal{E}_{\mathcal{F}}(C)$, this objective provides a more accurate characterization of the bias because it captures cancellation among residuals.
However, $\mathcal{G}(C)$ is tightly coupled to the directions, magnitudes, and distribution of the per-sample ZO gradients.
As a result, $\mathcal{E}_{\mathcal{G}}(C)$ does not yield a simple closed-form selection rule and is difficult to optimize under DP without resorting to costly search or task-specific assumptions.
These two formulations reveal the central tension.
The upper-bound objective $\mathcal{E}_{\mathcal{F}}(C)$ is easy to optimize but overly conservative.
The alignment-aware objective $\mathcal{E}_{\mathcal{G}}(C)$ is more accurate but difficult to optimize.
Since $\mathcal{F}(C)$ overestimates the bias, its induced threshold tends to be larger than what an alignment-aware objective would prefer.
Intuitively, we expect $C_{\mathcal{G}}^* \leq C_{\mathcal{F}}^*$, where $C_{\mathcal{G}}^*$ and $C_{\mathcal{F}}^*$ denote the optimal thresholds induced by $\mathcal{E}_{\mathcal{G}}(C)$ and $\mathcal{E}_{\mathcal{F}}(C)$, respectively.
Therefore, our goal is to find an intermediate formulation that is tighter than $\mathcal{F}(C)$, does not lead to an overly small threshold, and, crucially, still admits a closed-form optimization rule.

\noindent\textbf{Problem Statement.}
We formalize this goal with the following question: does there exist an error-objective formulation $\mathcal{E}_{\mathcal{R}}(C)$ with an induced optimal threshold $C_{\mathcal{R}}^*$ that satisfies all three of the following properties simultaneously?
\begin{enumerate}
    \item[i.] $\mathcal{E}_{\mathcal{R}}(C)$ should provide a tighter estimate than $\mathcal{E}_{\mathcal{F}}(C)$, so that it avoids the overly conservative threshold.
    \item[ii.] $\mathcal{E}_{\mathcal{R}}(C)$ should remain an upper bound on the true clipping bias at the minimizing threshold, thereby preventing the selection of an excessively small threshold.
    \item[iii.] $\mathcal{E}_{\mathcal{R}}(C)$ should yield an actionable rule, ideally reducing the problem to DP quantile estimation.
\end{enumerate}
    \subsection{Balancing Accuracy and Optimizability}
\label{sec14}
The previous section identifies the key requirement for private clipping-threshold selection: we need an error objective that is more accurate than the coarse upper bound $\mathcal{F}(C)$ but still simple enough to optimize without fine-tuning-based search.
In this section, we construct such an objective.
The main idea is to start from the alignment-aware objective $\mathcal{E}_{\mathcal{G}}(C)$ and derive a calibrated proxy objective that preserves part of its alignment information while retaining the quantile-based optimizability of $\mathcal{E}_{\mathcal{F}}(C)$.

We proceed in two stages.
First, for a fixed fine-tuning setting, we introduce a corrected bias formulation $r\cdot\mathcal{F}(C)$, where $r\in(0,1]$ captures the looseness of the triangle-inequality upper bound.
This immediately yields a closed-form quantile rule and avoids the overly conservative threshold induced by $\mathcal{F}(C)$.
However, the correction factor $r$ must be calibrated carefully: if $r$ is too small, the proxy may underestimate the true clipping bias.
We address this issue with self-consistent ladder calibration, giving a conservative threshold for a fixed setting (see \appendixref{app3} for the extended analysis).
Second, we remove setting- and task-specific calibration by defining a one-time offline calibration.
This produces the error objective used by our method.

\subsubsection{A Calibrated Proxy Objective for a Fixed Setting}
\label{sec15}

\noindent\emph{Step 1: Decomposing the alignment-aware objective.}
Recall that $\mathcal{F}(C)$ is easy to optimize but overly conservative, whereas $\mathcal{G}(C)$ is more accurate but difficult to optimize.
To bridge these two objectives, we rewrite $\mathcal{E}_{\mathcal{G}}(C)$ by factoring out $\mathcal{F}(C)$:
\begin{equation}
\label{eq6}
\begin{gathered}
\begin{aligned}
    \mathcal{E}_{\mathcal{G}}(C)
    = \mathcal{G}(C) + \lambda C
     &= \frac{\mathcal{G}(C)}{\mathcal{F}(C)} \cdot \mathcal{F}(C) + \lambda C\\
     &=\rho(C)\mathcal{F}(C)+\lambda C
\end{aligned}\\
    \frac{\partial}{\partial C}\mathcal{E}_{\mathcal{G}}(C)
    =\underbrace{\rho'(C)\cdot  \mathcal{F} (C)}_{\text{unknown}}
     + \underbrace{\rho(C)}_{\in [0,1]}
       \underbrace{\mathcal{F}(C)'+ \lambda}_{\text{closed form}}.
\end{gathered}
\end{equation}
Here, $\rho(C):=\mathcal{G}(C)/\mathcal{F}(C)$ measures how tight the triangle-inequality upper bound is at threshold $C$.
Whenever clipping occurs, $\rho(C)$ lies between $0$ and $1$, and both bounds are tight without further assumptions on the data.
This decomposition separates the optimizable parts, $\mathcal{F}'(C)$ and $\lambda$, from the difficult part, $\rho'(C)\mathcal{F}(C)$.

\noindent\emph{Step 2: Zeroth-order approximation of the alignment factor.}
The main obstacle in optimizing $\mathcal{E}_{\mathcal{G}}(C)$ is the dependence of $\rho(C)$ on $C$.
We therefore use a zeroth-order approximation of $\rho(C)$ with a constant correction factor $r \in (0,1]$.
This gives the proxy bias and the corresponding proxy error objective
\begin{equation}
    \label{eq7}
    \begin{aligned}
        \mathcal{R}(C) &:= r \cdot \mathcal{F}(C), \\
        \mathcal{E}_{\mathcal{R}}(C) &= \mathcal{R}(C) + \lambda C
        = r \cdot \mathcal{F}(C) + \lambda C, \quad r \in (0,1].
    \end{aligned}
\end{equation}
Intuitively, $r$ corrects the looseness of $\mathcal{F}(C)$ by accounting for the average cancellation among clipping residuals.
When $r=1$, $\mathcal{E}_{\mathcal{R}}(C)$ reduces to the coarse upper-bound objective $\mathcal{E}_{\mathcal{F}}(C)$.
When $r<1$, the proxy uses a tighter estimate of the clipping bias.
This objective remains easy to optimize, as its optimal threshold is again a quantile of gradient norms (\autoref{lem3}), which can be computed under DP using standard private quantile-selection mechanisms.

\begin{lemma}[Optimal Threshold via the Corrected Bias]
\label{lem3}
Consider a batch of size $b$ and a $K$-dimensional DPZO estimator.
For $r\in(0,1]$, the error objective $\mathcal{E}_{\mathcal{R}}(C)$ is minimized at
\begin{equation*}
    C_\mathcal{R}^* = \text{\emph{Quantile}}\left( \{\|g_x\|_2\}_{x \in B}, \, \left[1-\lambda/r\right]_{[0,1]} \right),
\end{equation*}
where $\left[\cdot\right]_{[0,1]} := \min\{1,\max\{0,\cdot\}\}$ and $\lambda := \sigma \sqrt{K} / b$.
\end{lemma}

Moreover, since $r\in(0,1]$, the induced quantile level $[1-\lambda/r]_{[0,1]}$ is no larger than $[1-\lambda]_{[0,1]}$, so $C_{\mathcal{R}}^* \leq C_{\mathcal{F}}^*$.
However, choosing $r$ too small may underestimate the true bias and select an overly small threshold.
Thus, $r$ must be calibrated to remain conservative while still tightening the upper bound $\mathcal{F}(C)$.

\noindent\emph{Step 3: Self-consistent calibration for a fixed setting.}
For simplicity, we first consider a fixed fine-tuning setting, $\phi \in \Phi$, where $\Phi$ denotes the set of practical hyperparameter settings.
Specifically, we partition the quantile space into $L$ ladders and treat $\rho(C)$ as a constant $\rho_\ell$ within each ladder.
The resulting ladderized error objective approximates the original one and takes the form in~\eqref{eq7} within each ladder.\footnote{As the number of ladders increases (larger $L$), the ladder width shrinks and the approximation error in the objective values vanishes.
For simplicity, we omit the term ``ladderized'' when referring to this objective.
See \appendixref{app3} for details.}
We then choose a self-consistent ladder whose proposed quantile lies roughly within that ladder.
The following informal theorem summarizes the guarantee, with the full statement deferred to \appendixref{app3}.
\begin{theorem}[Informal]
\label{thm1}
For a fixed setting $\phi\in \Phi$, let $r=\rho_{\mathrm{sc},\phi}$ denote the value selected by the \textit{self-consistent} (sc) ladder, and consider the error objective
\begin{equation*}
    \mathcal{E}_{\mathcal{R}}^{\phi}(C)=\rho_{\mathrm{sc},\phi}\cdot \mathcal{F}(C)+\lambda_\phi C.
\end{equation*}
If $\rho(C)$ is nondecreasing, the induced threshold upper bounds a near-optimal alignment-aware threshold.
\end{theorem}

The monotonicity assumption in \autoref{thm1} is well-motivated in DPZO.
As $C$ increases, fewer per-sample ZO gradients are clipped, leaving more structured and consistently aligned residuals with less cancellation, thereby increasing the alignment factor.
Empirical observations across multiple model--dataset combinations supporting the validity of this mild assumption are provided in \appendixref{app5}.

Recall that, in the exact objective $\mathcal{E}_{\mathcal{G}}(C)$ in \eqref{eq6}, the derivative includes the additional term $\rho'(C)\mathcal{F}(C)$.
Since $\rho'(C) \geq 0$ under our assumption, which is supported by empirical observations across a wide range of settings, and $\mathcal{F}(C) > 0$ whenever clipping occurs, this term is nonnegative and is omitted in our zeroth-order approximation.
At an interior critical point of the calibrated proxy, its derivative is zero.
Intuitively, adding back the nonnegative term favors an alignment-aware threshold to its left.
The preceding theorem formalizes this intuition: using self-consistent ladders, we calibrate the correction factor so that its induced threshold remains conservative relative to the alignment-aware objective.
For a fixed setting $\phi$, this yields a proxy with a bias estimate tighter than $\mathcal{F}(C)$, a conservative threshold relative to $\mathcal{G}(C)$, and a quantile-based solution.
In the next step, we replace this setting-specific calibration with a one-time offline calibration in the ``least noisy'' reference setting.

\subsubsection{One-Time Offline Calibration Across Settings and Tasks}
\label{sec16}

\noindent\emph{Step 4: Removing setting dependence.}
The setting-specific correction factor $\rho_{\mathrm{sc},\phi}$ satisfies the desired properties for a fixed setting $\phi$.
To cover all practical settings, we define the global correction as the largest factor across them:
\begin{equation*}
\begin{aligned}
    \mathcal{E}^{\mathrm{glob}}_{\mathcal{R}}(C)
    &:=
    r_{\mathrm{glob}}\cdot \mathcal{F}(C)+\lambda_\phi C,
    \quad
    r_{\mathrm{glob}} := \max_{\phi \in \Phi} \rho_{\mathrm{sc},\phi}.
\end{aligned}
\end{equation*}
By construction, $r_{\mathrm{glob}}\geq \rho_{\mathrm{sc},\phi}$ for every setting $\phi$.
Thus, the global objective is at least as conservative as the setting-specific self-consistent objective, while still tightening the coarse upper bound because $r_{\mathrm{glob}}\in(0,1]$.
To avoid exhaustive calibration over all hyperparameter settings $\phi=(K,b,\varepsilon,\delta)$, we note that the self-consistent ladder is primarily governed by $\lambda_\phi := \left(\frac{\sigma\sqrt{K}}{b}\right)_\phi$.
Using the Moments Accountant, we have
\begin{equation*}
\begin{aligned}
    \lambda_\phi
    \asymp
    \left(\frac{\sqrt{K\,T\,\log(1/\delta)}}{n\varepsilon}\right)_\phi
    \asymp
    \left(\frac{\sqrt{K\log(1/\delta)}}{\varepsilon}\right)_\phi,
\end{aligned}
\end{equation*}
where the second relation follows from the common practice of setting the number of steps so that $\sqrt{T}\propto n$~\cite{ref31,ref32,ref33}.
Thus, up to stable constants in the privacy accountant, $\lambda_\phi$ depends primarily on $(K,\varepsilon,\delta)$.

\autoref{cor1} (see \appendixref{app3}) shows that a larger $\lambda_\phi$ leads to a smaller selected correction factor.
The underlying intuition is that a larger $\lambda$ increases the noise cost of increasing $C$, thereby favoring a smaller clipping threshold.
This suggests using the factor from the least noisy setting for conservative global calibration:
\begin{equation*}
    r_{\mathrm{glob}} \gets \rho_{\mathrm{sc},\phi_{\min}},
    \qquad
    \phi_{\min} \in \arg\min_{\phi \in \Phi} \lambda_\phi.
\end{equation*}
In practice, we instantiate the reference setting with $(K,\varepsilon,\delta)=(16,6,10^{-5})$.
This configuration lies in a relatively low-$\lambda$ regime: $K=16$ is a moderate number of ZO directions, while $(\varepsilon,\delta)=(6,10^{-5})$ corresponds to a relatively loose privacy requirement.
We use this configuration as our ``least noisy'' reference setting in practice.

\noindent\emph{Step 5: Removing task dependence.}

\begin{figure}[htbp]
    \centering
    \includegraphics[width=0.65\columnwidth]{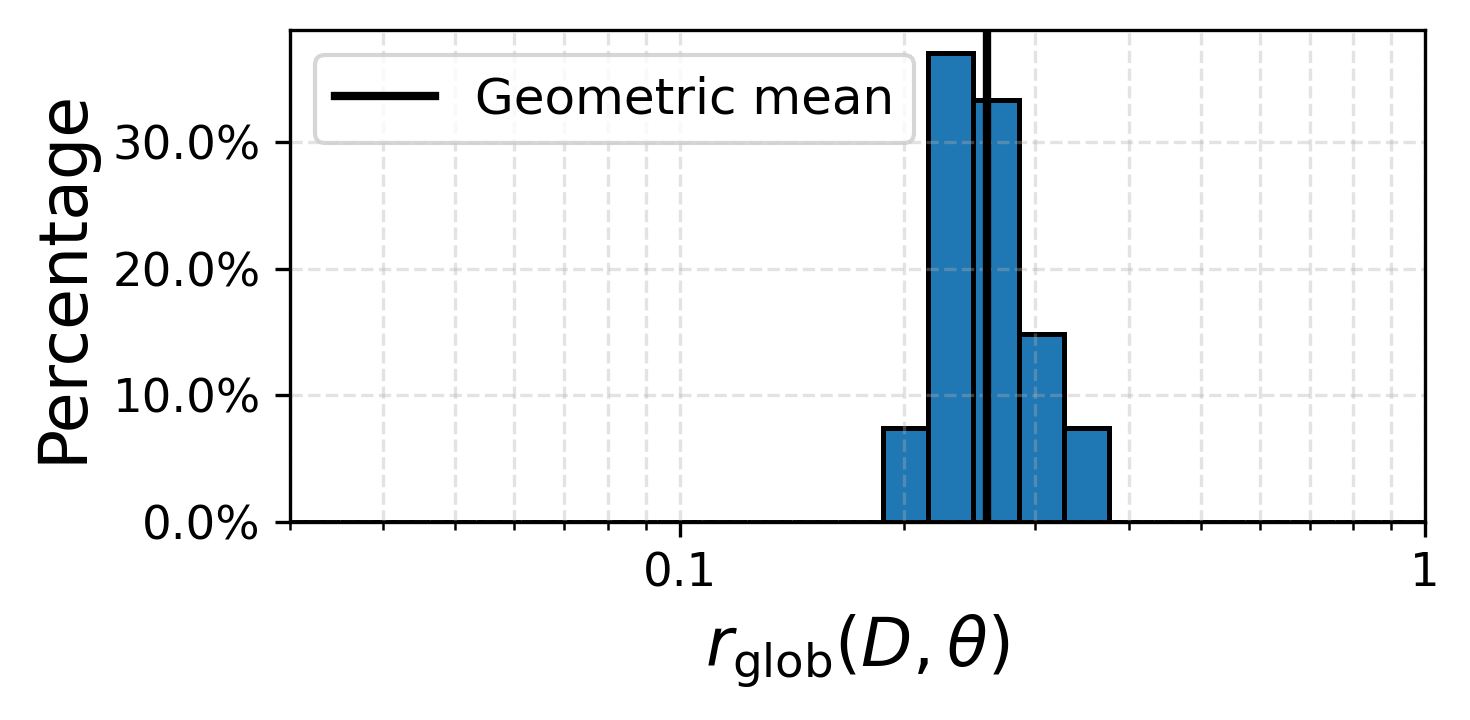}
    \caption{Concentration of the reference alignment factor across dataset--model combinations.}
    \label{fig2}
\end{figure}

We then evaluate the correction factor across dataset--model combinations and random seeds; experiment details are provided in \appendixref{app5}.
As shown in \autoref{fig2}, the resulting values are highly concentrated within the range $[0.213,0.345]$.
This concentration suggests that the correction factor is stable across tasks and does not require per-task calibration.
Since the correction factor is a ratio, we aggregate these values using the geometric mean and obtain $r_{\mathrm{ref}} = 0.258$.
The final error objective is therefore
\begin{equation*}
    \mathcal{E}_{\mathcal{R}}(C) = r_{\mathrm{ref}}\cdot \mathcal{F}(C) + \lambda C.
\end{equation*}
Roughly speaking, this calibration reflects the empirical observation that, in practice, the coarse upper-bound $\mathcal{F}(C)$ overestimates the bias by a factor of approximately $4\times$ or more due to spatial cancellation among the clipping residuals.
Calibrated once offline, our final objective avoids private search, tightens the coarse upper-bound objective, and reduces to DP quantile estimation.
It is designed to be conservative in general, with its practical effectiveness demonstrated through extensive experiments in \autoref{sec20}.

}

\section{\texttt{AutoClip-DPZO}}
\label{sec17}

{
\subsection{The Algorithm}
\label{sec18}

We now present our \texttt{AutoClip-DPZO} algorithm in \autoref{alg1}.
At a high level, it first uses a small portion of the privacy budget to select $C^*$, then uses the selected threshold for DPZO fine-tuning with the remaining budget.
More specifically, the target quantile is determined by \autoref{lem3}, where $r_{\mathrm{ref}}$ is calibrated offline.
We then integrate the private quantile mechanism~\cite{ref26} to select a private $C^*$ from per-sample ZO gradient norms at the initial model.
As a theory-guided practical optimization, we instead use a small predefined threshold $C=1$ when the full-budget target quantile is zero, skipping quantile estimation and saving its privacy budget for model fine-tuning.

\begin{algorithm}[htbp]
\caption{\texttt{AutoClip-DPZO}}
\label{alg1}
\LinesNumbered
\SetAlgoLined
\DontPrintSemicolon
\SetNoFillComment

\KwIn{Dataset $D$, initial model $\theta_{\mathrm{init}}$, number of directions $K$, batch size $b$, number of steps $T$, privacy budget $(\varepsilon,\delta)$, budget allocation $\alpha \in (0,1)$, upper bound $U$ on $C$, reference factor $r_{\mathrm{ref}} > 0$}
\KwOut{Privately fine-tuned model $\theta^*$}

{
$\sigma_{\mathrm{full}} \gets \mathsf{MomentAccount}(\varepsilon,\delta,b/|D|,T)$\;
\lIf{$\sigma_{\mathrm{full}}\sqrt{K} \ge b\,r_{\mathrm{ref}}$}{\tcp*[f]{Precheck}\newline
\hspace*{1em}\Return $\mathcal{M}_{\mathrm{DPZO}}\!\left(D,\theta_{\mathrm{init}},K,b,T,1,\varepsilon,\delta\right)$}
}

$\varepsilon_{\mathrm{search}} \gets \alpha\varepsilon,\quad \varepsilon_{\mathrm{release}} \gets (1-\alpha)\varepsilon$\;
{
$\sigma \gets \mathsf{MomentAccount}(\varepsilon_{\mathrm{release}},\delta,b/|D|,T)$\;
}
$q_{\mathrm{ref}} \gets \left[1-\frac{\sigma\sqrt{K}}{b\,r_{\mathrm{ref}}}\right]_{[0,1]}$\tcp*{Target quantile}
$D_{\mathrm{Est}} \gets \emptyset$\;
\For{$j = 1, \dots, \lceil |D|/b \rceil$}{
    $B_j \gets D[(j-1)b+1 : \min(jb, |D|)]$\;
    $D_{\mathrm{Est}} \gets D_{\mathrm{Est}} \cup \big\{\|\Delta_K(\theta_{\mathrm{init}},x)\|_2\big\}_{x \in B_j}$\;
}

\tcp{Stage 1: Private $C$ selection}
$C^* \gets \mathsf{Quantile}(D_{\mathrm{Est}}, 1/\sqrt{K}\cdot[1,U], \varepsilon_{\mathrm{search}}, q_{\mathrm{ref}})$\;

\tcp{Stage 2: Final DPZO fine-tuning}
$\theta^* \gets \mathcal{M}_{\mathrm{DPZO}}\!\left(D,\theta_{\mathrm{init}},K,b,T,C^*,\varepsilon_{\mathrm{release}} ,\delta\right)$\;

\Return $\theta^*$\;
\end{algorithm}

\noindent\textbf{Privacy guarantee.}
The early-return branch remains $(\varepsilon,\delta)$-DP since the precheck is public.
Otherwise, composing $\varepsilon_{\mathrm{search}}$-DP selection with $(\varepsilon_{\mathrm{release}},\delta)$-DP fine-tuning gives $(\varepsilon,\delta)$-DP because $\varepsilon_{\mathrm{search}}+\varepsilon_{\mathrm{release}}=\varepsilon$.
\subsection{Algorithm Analysis}
\label{sec19}

We now focus on the branch where quantile estimation is performed and analyze \texttt{AutoClip-DPZO} in terms of efficiency and hyperparameter robustness.
For memory usage, it preserves the memory advantage of ZO methods, since it introduces no backpropagation-based component and does not require a larger batch size.

\noindent\textbf{\texttt{AutoClip-DPZO} is computationally efficient.}
We normalize the runtime of one DPZO fine-tuning run with a predefined $C$ to one unit.
Under this normalization, \texttt{AutoClip-DPZO} adds only two components: constructing $D_{\mathrm{Est}}$ and running one private quantile query.
The quantile query is negligible, as it does not require GPU computation.
Constructing $D_{\mathrm{Est}}$ requires $|D|$ ZO-gradient computations, whereas the final fine-tuning stage requires $bT$ such computations.
Since fine-tuning typically iterates over the dataset multiple times, we have $|D|=o(bT)$, so the relative overhead is $\Theta(|D|/(bT))=o(1)$.
For example, when $|D|=10^3$, $b=64$, and $T=1,000$, constructing $D_{\mathrm{Est}}$ adds only about $1.6\%$ overhead.
By contrast, a search-based method that evaluates $M$ candidate thresholds requires $M$ search runs plus one final release run, yielding $\Theta(M+1)$ runtime.
Thus, \texttt{AutoClip-DPZO} avoids the linear search cost while incurring only negligible overhead over a single DPZO run.

\noindent\textbf{\texttt{AutoClip-DPZO} is robust to budget allocation.}
Unlike search-based methods, \texttt{AutoClip-DPZO} does not require tuning a search--release trade-off through repeated private fine-tuning.
The allocation factor $\alpha$ only needs to provide enough budget for private quantile estimation, and this requirement can be determined analytically.
For example, with domain bound $U=1024$, privacy budget $\varepsilon=6$, and dataset size $|D|=1024$, the high-utility guarantee of $\mathsf{Quantile}$ (see \appendixref{app9}) shows that any $\alpha \geq 6\%$ achieves a $10\%$ relative rank error.
Under a stricter budget $\varepsilon=3$, the same analysis gives $\alpha \geq 12\%$.
Therefore, a broad and predictable range of $\alpha$ values is sufficient, making the method insensitive to this hyperparameter.

\textit{Overall, \texttt{AutoClip-DPZO} selects $C$ automatically and privately, adds negligible overhead to a single DPZO run, and avoids fine-tuning-based hyperparameter search.}
}

\section{Experiments}
\label{sec20}

\subsection{Experimental Setup}
\label{sec21}
\noindent\textbf{Models, datasets, and metrics.}
We evaluate our method on two families of open-source pretrained models: the masked language model RoBERTa~\cite{ref34} and the autoregressive model OPT~\cite{ref35}, following prior work~\cite{ref17,ref28,ref16}.
To provide a broad evaluation, we consider model sizes ranging from 82M to 6.7B across these families.
For RoBERTa, we evaluate six classification tasks: SST-2 and SST-5 for sentiment analysis~\cite{ref36}, SNLI~\cite{ref37}, MNLI~\cite{ref38}, RTE for natural language inference~\cite{ref39}, and TREC for topic classification~\cite{ref40}.
For OPT, we evaluate two classification tasks, SST-2 and SST-5, and a generation task, SQuAD~\cite{ref41}.
We use the standard evaluation metrics used in prior studies~\cite{ref42,ref17}.

\noindent\textbf{Baselines.}
We compare \texttt{AutoClip-DPZO} with seven baselines.
For baselines that involve tunable hyperparameters, we report their best-performing configurations selected non-privately.
This strengthens the baselines and makes the comparison conservative for our method.

\textit{(i) Zero-Shot.}
This baseline tests the pretrained model directly and serves as a perfect-privacy lower bound.

\textit{(ii) Non-DP Optimal $C$.}
This baseline searches over candidate clipping thresholds and reports the best-performing model while using the full budget for each candidate.
It thus serves as an optimistic baseline that ignores the privacy cost of clipping selection.

\textit{(iii) Random Selection.}
This baseline randomly selects $C$ from the same candidate set as the non-DP optimal baseline.
We report the mean performance across all candidates.

\textit{(iv) Automatic Clipping~\cite{ref18}.}
Automatic Clipping is a widely used clipping selection method for DPSGD, in which gradients are normalized to have unit $\ell_2$ norm.
When adapted to DPZO, it still requires tuning a normalization factor $\gamma$.
Unlike clipping, normalization can enlarge gradients whose norms are smaller than one, and $\gamma$ controls this effect.
We report its best-performing configuration, giving this baseline an advantage.

\textit{(v) Strawman with DPZO ($C$).}
This baseline implements the two-stage search-and-release strategy discussed in \autoref{sec10}.
It first uses part of the privacy budget to select a clipping threshold through DPZO fine-tuning and then uses the remaining budget to train the final model.
We tune the budget-allocation factor $\alpha$ and report the best-performing result, giving this baseline a favorable setting.

\textit{(vi) Strawman with DPZO (Quantile).}
This baseline further replaces the candidate set of $C$ in baseline (v) with a set of candidate quantiles.
Unlike baseline (v), this allows the method to select a data-dependent $C$ outside the predefined grid, thereby potentially improving model utility.
Thus, it can be viewed as an improved version of the fixed quantile selection method in \autoref{tab3} from~\cite{ref43}.
Similarly, we tune $\alpha$ and report the best results to strengthen this baseline.

\textit{(vii) Upper-Bound Quantile (\autoref{lem2}).}
This baseline uses the quantile rule induced by the coarse bias upper bound in \autoref{lem2}.
It isolates the effect of our calibrated bias formulation by testing whether the standard triangle-inequality-based upper bound is sufficient for selecting $C$.

We do not include first-order adaptive clipping methods that rely on gradients from previous rounds.
Such recursive methods are difficult to apply reliably in the DPZO regime because the initial clipping threshold strongly affects the informativeness of the ZO-gradient signals.
In particular, initializing $C$ to a small value can over-clip the ZO gradients and prevent stable adaptive threshold updates.
We further discuss these methods in \appendixref{app1}.

\noindent\textbf{Privacy parameters and running settings.}
We fix $\delta=10^{-5}$, which is conservative for datasets of size $10^3$.
We vary $\varepsilon$ from $0.6$ to $6$ to evaluate \texttt{AutoClip-DPZO} under different privacy levels.
For baselines that search over $C$, we use a logarithmic grid with six candidates.
Using more candidates would increase both computational cost and privacy loss for search-based methods.
Unless otherwise specified, the candidate set is $1/\sqrt{K}\cdot\{4^{0}, 4^{1}, 4^{2}, 4^{3}, 4^{4}, 4^{5}\}$,
following the observation that the optimal threshold scales with $\sqrt{K}$~\cite{ref17}.
For \texttt{AutoClip-DPZO}, we use $1/\sqrt{K}\cdot[1,4^5]$ as the estimation range to conservatively cover possible gradient norms.

\noindent\textbf{Budget allocation.}
When the precheck in \autoref{alg1} does not skip estimation, we follow \autoref{sec19} and set $\alpha=6\%$ for $\varepsilon=6$ and $\alpha=12\%$ for $\varepsilon=3$, ensuring a $10\%$ relative rank error in private quantile selection.
For the strawman baselines (v) and (vi), $\alpha$ must be tuned because their search stages perform actual DP fine-tuning.
Therefore, we search over $\alpha \in \{6\%,12\%,25\%,50\%\}$ and report the best result.

\noindent\textbf{Learning parameters.}
For a fair comparison, we fix the batch size to $64$ and the number of fine-tuning steps to $1{,}000$.
For the main results in \autoref{tab2}, we set $K=16$ and specify other values of $K$ in the corresponding experiments.
For the learning rate, we report the best-performing model selected from the same search grid across all methods.
All experiments are conducted on an NVIDIA H200 GPU with 141GB of memory.
All experiments reported in our main results (\autoref{tab2}) are repeated three times with different random seeds, and we report the results in the form of $\mathrm{mean}\pm\mathrm{std}$.

Additional setup details are provided in \appendixref{app4}, with further results in \appendixref{app2}.

\newcommand{\result}[2]{\ensuremath{#1{\scriptstyle \pm #2}}}

\subsection{Main Results}
\begin{table*}[htbp]
\caption{\texttt{AutoClip-DPZO} utility versus baselines across tasks and privacy budgets. Dark/light gray denote best/second-best results. \textsuperscript{*} denotes results above Non-DP Optimal $C$.}
\label{tab2}
\centering
\renewcommand{\arraystretch}{1}
\setlength{\tabcolsep}{13pt}
\resizebox{\textwidth}{!}{
\begin{tabular}{c|c||cccccc}
\hline
\multirow{2}{*}{\textbf{Privacy}} &
\multirow{2}{*}{\textbf{Method}} &
\multicolumn{6}{c}{\textbf{RoBERTa-Large}} \\ \cline{3-8}
& &
\textbf{SST-2} &
\textbf{SST-5} &
\textbf{SNLI} &
\textbf{MNLI} &
\textbf{RTE} &
\textbf{TREC} \\
\hline
\hline

\multirow{6}{*}{$\varepsilon=6$}
& Random Selection
& \result{93.2}{0.3}
& \result{48.5}{0.5}
& \result{76.7}{1.0}
& \result{72.8}{0.9}
& \result{74.7}{0.2}
& \result{87.3}{0.5} \\
& Automatic Clipping~\cite{ref18}
& \result{93.2}{0.2}
& \result{44.1}{1.3}
& \result{79.2}{0.4}
& \result{74.8}{0.6}
& \result{74.1}{0.3}
& \result{88.6}{0.2} \\
& Two-Stage with DPZO ($C$)
& \cellcolor{gray!15}\result{93.5}{0.1}
& \result{48.6}{0.3}
& \cellcolor{gray!15}\result{79.9}{0.2}
& \cellcolor{gray!15}\result{75.0}{0.4}
& \result{75.5}{0.2}
& \cellcolor{gray!15}\result{88.8}{0.3} \\
& Two-Stage with DPZO (Quantile)
& \result{93.3}{0.2}
& \result{48.2}{0.2}
& \result{79.5}{0.3}
& \result{74.4}{0.3}
& \cellcolor{gray!15}\result{75.7}{0.4}
& \result{88.6}{0.5} \\
& Upper Bound (\autoref{lem2})
& \result{93.3}{0.1}
& \cellcolor{gray!15}\result{49.6}{0.5}
& \result{77.2}{0.5}
& \result{73.6}{1.4}
& \result{73.1}{0.5}
& \result{88.3}{0.3} \\
& \textbf{\texttt{AutoClip-DPZO} (Ours)}
& \cellcolor{gray!40}\result{93.8}{0.2}
& \cellcolor{gray!40}\result{50.4}{0.4}
& \cellcolor{gray!40}\result{80.2}{0.4}
& \cellcolor{gray!40}\result{75.2}{0.5}
& \cellcolor{gray!40}\result{76.2}{0.3}
& \cellcolor{gray!40}\result{90.4^{*}}{0.5} \\ \cline{2-8}
& Non-DP Optimal $C$
& \result{94.1}{0.2}
& \result{51.0}{0.5}
& \result{80.9}{0.4}
& \result{75.7}{0.3}
& \result{77.1}{0.2}
& \result{90.1}{0.5} \\
\hline

\multirow{6}{*}{$\varepsilon=3$}
& Random Selection
& \result{92.9}{0.1}
& \result{44.7}{0.4}
& \result{76.2}{0.5}
& \result{67.3}{5.1}
& \result{74.0}{0.5}
& \result{77.3}{0.8} \\
& Automatic Clipping~\cite{ref18}
& \result{93.0}{0.2}
& \result{41.8}{0.6}
& \cellcolor{gray!15}\result{79.0}{0.2}
& \cellcolor{gray!15}\result{71.2}{0.4}
& \result{72.2}{1.9}
& \cellcolor{gray!15}\result{87.5}{0.5} \\
& Two-Stage with DPZO ($C$)
& \result{92.7}{0.3}
& \result{47.4}{0.3}
& \result{78.8}{0.3}
& \result{69.8}{1.6}
& \cellcolor{gray!15}\result{74.3}{0.5}
& \result{85.2}{0.5} \\
& Two-Stage with DPZO (Quantile)
& \result{92.9}{0.1}
& \result{47.9}{0.3}
& \result{78.8}{0.2}
& \result{70.2}{0.9}
& \result{72.8}{0.6}
& \result{86.6}{0.3} \\
& Upper Bound (\autoref{lem2})
& \cellcolor{gray!15}\result{93.2}{0.1}
& \cellcolor{gray!15}\result{48.0}{0.3}
& \result{74.1}{0.9}
& \result{66.9}{3.0}
& \result{70.9}{0.2}
& \result{87.3}{0.2} \\
& \textbf{\texttt{AutoClip-DPZO} (Ours)}
& \cellcolor{gray!40}\result{93.3}{0.1}
& \cellcolor{gray!40}\result{48.6}{0.4}
& \cellcolor{gray!40}\result{79.5}{0.2}
& \cellcolor{gray!40}\result{71.9}{0.5}
& \cellcolor{gray!40}\result{75.9}{0.2}
& \cellcolor{gray!40}\result{88.7}{0.3} \\ \cline{2-8}
& Non-DP Optimal $C$
& \result{93.9}{0.3}
& \result{49.1}{1.1}
& \result{79.9}{0.2}
& \result{73.1}{0.5}
& \result{76.5}{0.8}
& \result{89.7}{0.4} \\
\hline

\multicolumn{1}{c|}{Perfect} &
\multicolumn{1}{c||}{Zero-Shot}
& \result{66.0}{9.1}
& \result{32.7}{3.9}
& \result{42.7}{7.1}
& \result{45.8}{6.6}
& \result{50.6}{2.8}
& \result{30.4}{5.4} \\
\hline
\hline

\multicolumn{2}{c||}{} &
\multicolumn{6}{c}{\textbf{OPT-350M}} \\ \cline{3-8}
\multicolumn{2}{c||}{} &
\multicolumn{2}{c|}{\textbf{SST-2}} &
\multicolumn{2}{c|}{\textbf{SST-5}} &
\multicolumn{2}{c}{\textbf{SQuAD}} \\
\hline

\multirow{6}{*}{$\varepsilon=6$}
& Random Selection
& \multicolumn{2}{c|}{\result{87.3}{0.2}}
& \multicolumn{2}{c|}{\result{65.2}{0.5}}
& \multicolumn{2}{c}{\result{53.2}{0.7}} \\
& Automatic Clipping~\cite{ref18}
& \multicolumn{2}{c|}{\result{87.5}{0.1}}
& \multicolumn{2}{c|}{\result{62.8}{1.4}}
& \multicolumn{2}{c}{\cellcolor{gray!15}\result{54.8}{0.2}} \\
& Two-Stage with DPZO ($C$)
& \multicolumn{2}{c|}{\result{87.8}{0.3}}
& \multicolumn{2}{c|}{\result{66.7}{0.4}}
& \multicolumn{2}{c}{\result{53.5}{0.5}} \\
& Two-Stage with DPZO (Quantile)
& \multicolumn{2}{c|}{\result{87.2}{0.3}}
& \multicolumn{2}{c|}{\result{67.3}{0.8}}
& \multicolumn{2}{c}{\result{53.7}{0.5}} \\
& Upper Bound (\autoref{lem2})
& \multicolumn{2}{c|}{\cellcolor{gray!15}\result{88.1}{0.2}}
& \multicolumn{2}{c|}{\cellcolor{gray!15}\result{68.2}{0.2}}
& \multicolumn{2}{c}{\result{54.6}{0.3}} \\
& \textbf{\texttt{AutoClip-DPZO} (Ours)}
& \multicolumn{2}{c|}{\cellcolor{gray!40}\result{88.5}{0.3}}
& \multicolumn{2}{c|}{\cellcolor{gray!40}\result{68.7}{0.2}}
& \multicolumn{2}{c}{\cellcolor{gray!40}\result{55.8^{*}}{0.3}} \\ \cline{2-8}
& Non-DP Optimal $C$
& \multicolumn{2}{c|}{\result{88.8}{0.1}}
& \multicolumn{2}{c|}{\result{69.2}{0.4}}
& \multicolumn{2}{c}{\result{55.6}{0.5}} \\
\hline

\multirow{6}{*}{$\varepsilon=3$}
& Random Selection
& \multicolumn{2}{c|}{\result{86.6}{0.2}}
& \multicolumn{2}{c|}{\result{63.1}{0.4}}
& \multicolumn{2}{c}{\result{51.7}{0.4}} \\
& Automatic Clipping~\cite{ref18}
& \multicolumn{2}{c|}{\result{86.2}{0.1}}
& \multicolumn{2}{c|}{\result{59.9}{2.3}}
& \multicolumn{2}{c}{\result{50.2}{0.2}} \\
& Two-Stage with DPZO ($C$)
& \multicolumn{2}{c|}{\result{87.5}{0.2}}
& \multicolumn{2}{c|}{\cellcolor{gray!15}\result{66.0}{0.6}}
& \multicolumn{2}{c}{\result{50.7}{0.4}} \\
& Two-Stage with DPZO (Quantile)
& \multicolumn{2}{c|}{\result{87.0}{0.1}}
& \multicolumn{2}{c|}{\result{65.9}{0.5}}
& \multicolumn{2}{c}{\result{53.3}{0.2}} \\
& Upper Bound (\autoref{lem2})
& \multicolumn{2}{c|}{\cellcolor{gray!15}\result{87.8}{0.3}}
& \multicolumn{2}{c|}{\result{65.7}{0.2}}
& \multicolumn{2}{c}{\cellcolor{gray!15}\result{53.8}{0.5}} \\
& \textbf{\texttt{AutoClip-DPZO} (Ours)}
& \multicolumn{2}{c|}{\cellcolor{gray!40}\result{88.3}{0.2}}
& \multicolumn{2}{c|}{\cellcolor{gray!40}\result{67.8}{0.5}}
& \multicolumn{2}{c}{\cellcolor{gray!40}\result{54.6}{0.2}} \\ \cline{2-8}
& Non-DP Optimal $C$
& \multicolumn{2}{c|}{\result{88.5}{0.3}}
& \multicolumn{2}{c|}{\result{68.2}{0.2}}
& \multicolumn{2}{c}{\result{55.0}{0.2}} \\
\hline

\multicolumn{1}{c|}{Perfect} &
\multicolumn{1}{c||}{Zero-Shot}
& \multicolumn{2}{c|}{\result{60.1}{4.5}}
& \multicolumn{2}{c|}{\result{43.2}{5.6}}
& \multicolumn{2}{c}{\result{25.4}{8.3}} \\
\hline
\end{tabular}
}
\end{table*}

\noindent\textbf{\texttt{AutoClip-DPZO} provides strong model utility.}
As shown in \autoref{tab2}, \texttt{AutoClip-DPZO} consistently outperforms all baselines, improving over the best baseline for each setting by up to $1.8$ points.
This is notable because the baselines benefit from additional non-DP hyperparameter tuning, as described in \autoref{sec21}.
Moreover, across tasks and settings, the baselines exhibit substantially less stable utility.
Specifically, Automatic Clipping, the two-stage protocol with $C$, the two-stage protocol with quantile, and the upper-bound quantile method can suffer performance drops of up to $7.9$, $3.9$, $3.1$, and $5.4$ points, respectively.
This suggests that these methods do not generalize well.
By contrast, \texttt{AutoClip-DPZO} achieves utility comparable to that of the non-DP optimal-$C$ baseline obtained by search, and even outperforms it in two cases.
One reason is that our quantile-based rule can select $C$ at a finer granularity, whereas search-based methods are restricted to a predefined grid.
At the same time, increasing grid resolution would further increase runtime and, for private search methods, privacy costs.
Indeed, the $C$ selected by \texttt{AutoClip-DPZO} often lies between search candidates, as shown in \autoref{tab11}.
Moreover, the zero-shot baseline consistently yields low utility, confirming the need for private fine-tuning.
Random $C$ selection also performs poorly across settings, highlighting the need for principled selection.

\noindent\textbf{Fixed clipping strategies generalize poorly.}
\autoref{tab3} shows that no single fixed strategy, whether a fixed quantile in $\{0.25, 0.5, 0.75\}$ or a fixed clipping threshold, performs well across tasks and settings.\footnote{The $50\%$ quantile is the median tracked in~\cite{ref43}.
However, repeated estimation consumes additional privacy budget without improving utility (\autoref{tab9}), so we perform quantile estimation only once for all methods.}
Indeed, each clipping threshold or quantile is optimal in only a few cases and can perform poorly in others.
Therefore, using a single fixed approach does not generalize well across settings and is thus highly unreliable.
By contrast, across the $12$ settings in \autoref{tab3}, our method achieves an average utility improvement of $0.22$ points over the best fixed strategy in each setting.
This observation further highlights the need for a principled clipping-threshold selection method.
Results at $K=64$ are provided in~\appendixref{app2}.

\begin{table}[!htbp]
\caption{Fixed clipping strategies do not generalize across settings ($K=16$).}
\centering
\small
\resizebox{1\linewidth}{!}{
\begin{tabular}{c|cc||cc|cc}
\hline
\multirow{2}{*}{\textbf{Privacy}} &\multicolumn{2}{c||}{\multirow{2}{*}{
\textbf{
Method
}}}  & \multicolumn{2}{c}{\textbf{OPT-350M}}
& \multicolumn{2}{c}{\textbf{RoBERTa-Large}}  \\
\cline{4-5} \cline{6-7}
& & &\multicolumn{1}{c}{\textbf{SST-2}} &\multicolumn{1}{c}{\textbf{SST-5}} & \textbf{SST-2} &\textbf{SST-5} \\
\hline \hline
& &\multicolumn{4}{c}{--- $K=16$ ---} \\ \cline{2-7}
\multirow{7}{*}{$\varepsilon=6$}
&\multirow{3}{*}{\textbf{Fixed Value}}&\multicolumn{1}{|c||}{$C=16$}  &\cellcolor{gray!40}88.8  &67.5  &\cellcolor{gray!15}93.4  &49.0   \\
& &\multicolumn{1}{|c||}{$C=64$} &88.4  &66.4  &91.6  &50.2    \\
& &\multicolumn{1}{|c||}{$C=128$}  &86.5  &65.7  &90.7  &46.4   \\ \cline{2-7}
& \multirow{3}{*}{\textbf{Fixed Quantile}}  &\multicolumn{1}{|c||}{$Q=0.25$}  &88.0 &\cellcolor{gray!40}69.0 &93.1 &\cellcolor{gray!40}51.1    \\
& &\multicolumn{1}{|c||}{$Q=0.50$}  &88.1  &68.5  &92.9  &49.0    \\
& &\multicolumn{1}{|c||}{$Q=0.75$}  &\cellcolor{gray!15}88.5  &\cellcolor{gray!40}69.0  &92.9  &50.2     \\  \cline{2-7}
&\multicolumn{2}{c||}{\textbf{\texttt{AutoClip-DPZO}}}  &\multicolumn{1}{c}{\cellcolor{gray!15}88.5}
&\multicolumn{1}{c|}{\cellcolor{gray!15}68.7}  & \cellcolor{gray!40}93.8
& \cellcolor{gray!15}50.4 \\  \hline \hline
\multirow{7}{*}{$\varepsilon=3$}
&\multirow{3}{*}{\textbf{Fixed Value}}&\multicolumn{1}{|c||}{$C=16$}  &\cellcolor{gray!15}88.5   &\cellcolor{gray!15}68.1  &92.7  &44.5   \\
& &\multicolumn{1}{|c||}{$C=64$} &86.8   &64.9  &91.1  &48.1     \\
& &\multicolumn{1}{|c||}{$C=128$}  &85.8   &63.8  &90.4  &46.1   \\ \cline{2-7}
& \multirow{3}{*}{\textbf{Fixed Quantile}}  &\multicolumn{1}{|c||}{$Q=0.25$}  &\cellcolor{gray!40}88.7  &67.0 &\cellcolor{gray!15}92.9  &48.3    \\
& &\multicolumn{1}{|c||}{$Q=0.50$}  &88.3  &\cellcolor{gray!40}68.7  &92.7 &\cellcolor{gray!40}48.7      \\
& &\multicolumn{1}{|c||}{$Q=0.75$}  &87.7  &66.0 &92.7  &48.1   \\  \cline{2-7}
&\multicolumn{2}{c||}{\textbf{\texttt{AutoClip-DPZO}}}  &\multicolumn{1}{c}{88.3}
&\multicolumn{1}{c|}{67.8}&\cellcolor{gray!40}93.3
& \cellcolor{gray!15}48.6\\
\hline \hline
\multirow{7}{*}{$\varepsilon=1$}
&\multirow{3}{*}{\textbf{Fixed Value}}&\multicolumn{1}{|c||}{$C=16$}  &85.1   &59.5  &\cellcolor{gray!15}88.9  &47.0   \\
& &\multicolumn{1}{|c||}{$C=64$} &81.9   &56.9  &87.6  &44.8     \\
& &\multicolumn{1}{|c||}{$C=128$}  &73.9   &40.7  &83.7  &40.8   \\ \cline{2-7}
& \multirow{3}{*}{\textbf{Fixed Quantile}}  &\multicolumn{1}{|c||}{$Q=0.25$}  &\cellcolor{gray!15}85.2  &59.9 &\cellcolor{gray!15}88.9  &\cellcolor{gray!15}47.1    \\
& &\multicolumn{1}{|c||}{$Q=0.50$}  &85.0  &\cellcolor{gray!40}60.4  &88.2  &46.5      \\
& &\multicolumn{1}{|c||}{$Q=0.75$}  &83.6  &\cellcolor{gray!40}60.4  &87.8  &45.5   \\  \cline{2-7}
&\multicolumn{2}{c||}{\textbf{\texttt{AutoClip-DPZO}}}  &\multicolumn{1}{c}{\cellcolor{gray!40}86.2}
&\multicolumn{1}{c|}{\cellcolor{gray!15}59.9}&\cellcolor{gray!40}92.1
& \cellcolor{gray!40}47.9\\
\hline
\end{tabular}
}
\label{tab3}
\end{table}

\noindent\textbf{\texttt{AutoClip-DPZO} is highly efficient.}
The runtime results in \autoref{tab4} are consistent with the analysis in \autoref{sec19}.
When the runtime of DPZO fine-tuning with a predefined $C$ is normalized to one unit, \texttt{AutoClip-DPZO} incurs only $3\%$--$27\%$ additional overhead.
This small overhead arises from constructing the estimation dataset and performing a single private quantile query, without requiring any additional fine-tuning runs.
In contrast, the two-stage search baseline incurs a constant-factor overhead of up to $7.87\times$ because it must fine-tune once for each candidate $C$ before the final release run.
As a result, \texttt{AutoClip-DPZO} achieves a speedup of up to $6.90\times$ over this search-based baseline.
In wall-clock time, searching over $C$ can take nearly $10$ days on an NVIDIA H200 when $K=64$ for the 6.7B model.
Since the cost of DPZO scales roughly linearly with $K$, extending such a search to $K=256$, a setting used to improve performance in prior work~\cite{ref17}, would require more than a month.
This makes exhaustive or grid-based search impractical for resource-constrained data owners.
By avoiding repeated fine-tuning, \texttt{AutoClip-DPZO} preserves the memory efficiency of DPZO while substantially reducing its runtime overhead, making it more practical.

\begin{table}[htbp]
\caption{OPT runtime comparison ($K=64$).}
\centering
\resizebox{0.90\linewidth}{!}{
\renewcommand{\arraystretch}{1}
\setlength{\tabcolsep}{6pt}
\begin{tabular}{c|c||cc}
\hline
\multirow{2}{*}{\textbf{Dataset}} &
\multirow{2}{*}{\begin{tabular}[c]{@{}c@{}}\textbf{Model} \\ \textbf{Size}\end{tabular}} &
\multicolumn{2}{c}{\textbf{Hardware Time (hours)}} \\ \cline{3-4}
& & \textbf{Two-Stage} & \textbf{\texttt{AutoClip-DPZO}} \\
\hline
\hline
\multirow{4}{*}{\textbf{SST-2}}
& \textbf{125M} & 4.87h~($7.74\times$) & 0.72h~($1.14\times$) \\
& \textbf{350M} & 8.97h~($7.13\times$) & 1.30h~($1.03\times$) \\
& \textbf{1.3B} & 19.13h~($7.46\times$) & 2.80h~($1.09\times$) \\
& \textbf{6.7B} & 75.50h~($7.28\times$) & 11.12h~($1.07\times$) \\
\hline
\multirow{4}{*}{\textbf{SQuAD}}
& \textbf{125M} & 28.15h~($7.55\times$) & 4.13h~($1.11\times$) \\
& \textbf{350M} & 39.42h~($7.41\times$) & 6.22h~($1.17\times$) \\
& \textbf{1.3B} & 71.07h~($7.53\times$) & 11.19h~($1.19\times$) \\
& \textbf{6.7B} & 238.96h~($7.87\times$) & 38.63h~($1.27\times$) \\
\hline
\end{tabular}
}
\label{tab4}
\end{table}

\begin{table}[htbp]
\caption{RoBERTa-Large utility at small privacy budgets ($K=16$).}
\centering
\resizebox{0.90\linewidth}{!}{
\renewcommand{\arraystretch}{1}
\setlength{\tabcolsep}{12pt}
\begin{tabular}{c|c||cc}
\hline
\textbf{Privacy} & \textbf{Method} & \textbf{SST-2} & \textbf{SNLI} \\
\hline
\hline
\multirow{3}{*}{$\varepsilon=0.6$}
& Upper Bound & 90.3 & 64.6 \\
& Automatic Clipping & 90.5 & 65.9 \\
& \textbf{\texttt{AutoClip-DPZO}} & \cellcolor{gray!40}91.1 & \cellcolor{gray!40}67.8 \\
\cline{1-4}
\multirow{3}{*}{$\varepsilon=1$}
& Upper Bound & 91.5 & 71.9 \\
& Automatic Clipping & 91.5 & 73.7 \\
& \textbf{\texttt{AutoClip-DPZO}} & \cellcolor{gray!40}92.1 & \cellcolor{gray!40}74.7 \\
\hline
\end{tabular}
}
\label{tab5}
\end{table}

\begin{table}[htbp]
\caption{RoBERTa-Large utility across $K$ ($\varepsilon=6$). \textsuperscript{*} denotes results above Non-DP Optimal $C$.}
\centering
\resizebox{0.90\linewidth}{!}{
\renewcommand{\arraystretch}{1}
\setlength{\tabcolsep}{12pt}
\begin{tabular}{c|c||cc}
\hline
\textbf{$K$} & \textbf{Method} & \textbf{SST-2} & \textbf{SST-5} \\
\hline
\hline
\multirow{3}{*}{$32$}
& Upper Bound & 93.6 & 48.3 \\
& \textbf{\texttt{AutoClip-DPZO}} & \cellcolor{gray!40}93.7 & \cellcolor{gray!40}50.8 \\
\cline{2-4}
& Non-DP Optimal $C$ &93.7  &51.1 \\
\cline{1-4}
\multirow{3}{*}{$64$}
& Upper Bound & 93.5 & 50.5 \\
& \textbf{\texttt{AutoClip-DPZO}} & \cellcolor{gray!40}93.9 & \cellcolor{gray!40}51.0 \\
\cline{2-4}
& Non-DP Optimal $C$ & 94.2 & 51.5 \\
\cline{1-4}
\multirow{3}{*}{$128$}
& Upper Bound & 93.9 & 51.2 \\
& \textbf{\texttt{AutoClip-DPZO}} & \cellcolor{gray!40}94.2 & \cellcolor{gray!40}{$52.3^{*}$} \\ \cline{2-4}
& Non-DP Optimal $C$ & 94.4 & 52.2 \\
\hline
\end{tabular}
}
\label{tab6}
\end{table}

{
\clubpenalty=10000
\noindent\textbf{\texttt{AutoClip-DPZO} generalizes across privacy budgets and numbers of directions.}
\autoref{tab5} and \autoref{tab6} evaluate \texttt{AutoClip-DPZO} under smaller privacy budgets and larger values of $K$.
In small privacy-budget settings, the precheck in \autoref{alg1} allows \texttt{AutoClip-DPZO} to use $C=1$ directly, avoiding private estimation and saving privacy budget.
However, without the correction factor, the upper-bound rule is too loose to trigger this shortcut.
For the upper-bound method, we relax the relative rank-error guarantee from $10\%$ to $20\%$, yielding $\alpha=30\%$ for $\varepsilon=0.6$ and $\alpha=18\%$ for $\varepsilon=1$, thereby reserving more privacy budget for final fine-tuning.
The two-stage protocol is less suitable in this regime because its search stage requires DP fine-tuning and becomes both privacy- and runtime-expensive, especially when $K$ is large.
Thus, we primarily compare against the upper-bound baseline and additionally consider Automatic Clipping in the low-privacy-budget regime, where our predicted zeroth quantile suggests that this approach may perform well.
}

Across these experiments, \texttt{AutoClip-DPZO} consistently outperforms the baselines, improving upon the best-performing one in each setting by up to $2.5$ points.
This supports our analysis in \autoref{sec16}: the calibrated objective $\mathcal{E}_{\mathcal{R}}(C)$ characterizes the error more accurately than $\mathcal{E}_{\mathcal{F}}(C)$.
Surprisingly, in the noisy regime, our method even outperforms Automatic Clipping, demonstrating the effectiveness of our practical optimization.
Thus, our calibrated quantile rule transfers well across privacy budgets, tasks, and numbers of directions.
Additional results on distilled and larger models are provided in \autoref{tab13}.

\subsection{Ablation Study}

\noindent\textbf{Impact of $\alpha$.}
\autoref{tab7} shows that the theoretically guided choice of $\alpha$, which guarantees a $10\%$ relative rank error in private quantile selection, is sufficient to achieve competitive utility in our experiments.
Thus, $\alpha$ does not require tuning.
Notably, the corresponding search budget is only $0.36$, showing our method requires only a small budget for clipping selection.
This contrasts with search-based methods, whose search stages perform fine-tuning and can become noisy and thus unreliable under small budgets.

\begin{table}[ht]
\caption{RoBERTa-Large utility across $\alpha$ and privacy budgets ($K=16$).}
\centering
\resizebox{0.90\linewidth}{!}{
\renewcommand{\arraystretch}{1}
\setlength{\tabcolsep}{12pt}
\begin{tabular}{c|c||ccc}
\hline
\multirow{2}{*}{\textbf{Privacy}} &
\multirow{2}{*}{\textbf{Dataset}} &
\multicolumn{3}{c}{\textbf{$\varepsilon_{\mathrm{search}}=\alpha\varepsilon$}} \\
\cline{3-5}
& & $0.18$ & $0.36$ & $0.72$ \\
\hline
\hline
\multirow{2}{*}{$\varepsilon=6$}
& \textbf{SNLI} & 80.1 & \cellcolor{gray!15}80.2 & \cellcolor{gray!40}80.5 \\
& \textbf{SST-5} & 49.6 & \cellcolor{gray!40}50.4 & \cellcolor{gray!15}50.2 \\
\cline{1-5}
\multirow{2}{*}{$\varepsilon=3$}
& \textbf{SNLI} & \cellcolor{gray!15}79.5 & \cellcolor{gray!40}79.6 & 78.4 \\
& \textbf{SST-5} & \cellcolor{gray!40}48.6 & \cellcolor{gray!40}48.6 & 48.0 \\
\hline
\end{tabular}
}
\label{tab7}
\end{table}

\noindent\textbf{Impact of the calibration method for $r_{\mathrm{ref}}$.}
As discussed in \autoref{sec16}, $r_{\mathrm{ref}}$ captures the ratio between the actual clipping bias and its upper bound.
The geometric mean across model--dataset combinations therefore provides a natural calibration for this ratio.
\autoref{tab8} shows that using the minimum or maximum value of $r_{\mathrm{ref}}$ can improve utility in some cases, but these choices are less stable than the geometric mean, which consistently achieves the best or second-best utility.
Even under the two extreme calibrations, $75\%$ of the cases still outperform or match the best baseline in \autoref{tab2}.
This suggests that \texttt{AutoClip-DPZO} is relatively insensitive to the exact calibration choice.
Because the calibrated alignment factors are concentrated across models and datasets, we further show in \autoref{tab15} that alternative choices, such as the median or arithmetic mean, also achieve better model utility than the baselines.

\begin{table}[ht]
\caption{Utility across $r_{\mathrm{ref}}$ calibrations ($\varepsilon=6,\ K=16$). \textsuperscript{*} denotes results above the best baseline in~\autoref{tab2}.}
\centering
\resizebox{1\linewidth}{!}{
\renewcommand{\arraystretch}{1}
\setlength{\tabcolsep}{12pt}
\begin{tabular}{c|c||ccc}
\hline
\multirow{2}{*}{\textbf{Model}} &
\multirow{2}{*}{\textbf{Dataset}} &
\multicolumn{3}{c}{\textbf{$r_{\mathrm{ref}}$}} \\
\cline{3-5}
& & $\min$ & $\mathrm{GM}$ & $\max$ \\
\hline
\hline
\multirow{2}{*}{\textbf{RoBERTa-Large}}
& \textbf{SST-2} & 93.2 & \cellcolor{gray!40}{$93.8^{*}$} & {$93.6^{*}$} \\
& \textbf{SST-5} & \cellcolor{gray!40}{$50.7^{*}$} & \cellcolor{gray!15}{$50.4^{*}$} & 49.5 \\
\cline{1-5}
\multirow{2}{*}{\textbf{OPT-350M}}
& \textbf{SST-2} & {$88.5^{*}$} & \cellcolor{gray!15}{$88.5^{*}$} & \cellcolor{gray!40}{$88.9^{*}$} \\
& \textbf{SST-5} & \cellcolor{gray!40}{$69.0^{*}$} & \cellcolor{gray!15}{$68.7^{*}$} & {$68.5^{*}$} \\
\hline
\end{tabular}
}
\label{tab8}
\end{table}

\section{Conclusion}
\label{sec22}

In this work, we study clipping selection for private zeroth-order LLM fine-tuning.
While DPZO reduces memory overhead by avoiding backpropagation, its utility is highly sensitive to the clipping threshold, and existing fine-tuning-based search is computationally expensive and privacy-inefficient.
To address this, we propose \texttt{AutoClip-DPZO}, a practical, fine-tuning-free method for automatically selecting $C$ under DP.
By analyzing the noise--bias trade-off in DPZO, we develop a calibrated proxy objective that is tighter than standard worst-case bias bounds and reduces threshold selection to efficient private quantile estimation.
Extensive experiments show that \texttt{AutoClip-DPZO} achieves utility close to non-DP greedy search, consistently outperforms baselines, and incurs negligible additional overhead.
Promising directions for future research include extending our framework to other private ZO optimizers, adaptive noise schedules, and parameter-efficient fine-tuning methods.
While \texttt{AutoClip-DPZO} reduces the cost of threshold selection, DPZO still trades memory efficiency for additional forward passes, motivating methods that jointly improve memory efficiency, runtime, and utility.
\texttt{AutoClip-DPZO} should naturally extend to such methods, provided that the required $K$ remains in a similar regime and induces comparable noise levels.

}

\appendices
\section{Related Work}
\label{app1}

ZO fine-tuning replaces backpropagation with memory-efficient forward-only evaluations along random directions~\cite{ref28}.
DPZO extends this approach to mitigate membership inference~\cite{ref4,ref5,ref6} and data-extraction attacks~\cite{ref7,ref8}, while injecting noise in a lower-dimensional space~\cite{ref16,ref15,ref17,ref14}.
Multi-direction variants reduce the variance of earlier single-direction DPZO~\cite{ref16,ref15,ref14,ref17} and can outperform DP-AdamW~\cite{ref17,ref12}.
Prior work shows that DPZO requires larger and more variable clipping thresholds $C$ than first-order methods, yet typically selects $C$ through repeated fine-tuning on private data~\cite{ref17,ref16,ref14,ref15}.
This makes clipping-threshold selection particularly challenging due to its high computational and privacy costs.

We next explain why existing first-order methods for selecting $C$ do not apply to DPZO.
\emph{a.~Fixed $C$ or fixed quantile:}
first-order methods commonly use a small fixed $C$~\cite{ref44,ref45,ref46} or a fixed quantile of gradient norms~\cite{ref43}, since high-dimensional noise can dominate clipping bias.
However, these approaches perform poorly across DPZO settings (\autoref{tab3}), where clipping bias also becomes important.
\emph{b.~Normalization-based methods:}
some DPSGD methods normalize gradients to a constant (e.g., $C=1$) and correct the resulting bias with extra terms~\cite{ref18,ref31}, but these correction terms become sensitive under DPZO's noise regime and perform poorly in our experiments (\autoref{sec20}).
\emph{c.~Recursive methods:} these methods update $C$ using historical gradients~\cite{ref47,ref43,ref32,ref48,ref49}, but DPZO lacks a reliable small-$C$ initialization across settings, while tuning initialization incurs additional privacy costs.
\emph{Public-data-based methods.} Several methods leverage auxiliary public data~\cite{ref50,ref51,ref52,ref53,ref54,ref55}.
However, suitable public data may be unavailable, and access to similar public data can increase membership-inference risks~\cite{ref56,ref57}.
More broadly, existing first-order analyses formulate the noise--bias trade-off using either loose upper bounds~\cite{ref32} or exact data-dependent bias~\cite{ref44}.
Although prior work has shown that geometric cancellation can favor smaller $C$~\cite{ref44,ref45,ref58}, these formulations agree in the noise-dominated first-order regime, which does not hold in DPZO.
In addition, tighter privacy accounting alone remains insufficient for efficient hyperparameter tuning.
Existing privacy bounds are essentially tight under $(\varepsilon,\delta)$-DP and RDP~\cite{ref59}, while searching multiple candidates still incurs a constant-factor increase in privacy costs~\cite{ref60,ref61} and computational overhead~\cite{ref60,ref62}.
In contrast, our method avoids repeated fine-tuning and uses as little as $6\%$ of the total budget for selection.
Another thread of research focuses on whether ZO inherently amplifies privacy due to the randomness of the direction.
Unfortunately, this randomness alone provides no inherent privacy advantage~\cite{ref29}.
Tighter privacy bounds for modified DPZO with orthonormal directions require restrictive parameter regimes and strong assumptions on the loss functions~\cite{ref30}, whereas our approach does not rely on such assumptions.

To the best of our knowledge, our work is the first to address the challenges posed by DPZO's distinct noise regime with a principled, efficient, and automatic selection method that requires no additional hyperparameter tuning.

\section{Extended Experimental Evaluation}
\label{app2}

We present additional experiments covering estimation schedules, different model sizes, algorithm runtime, varying $K$ across models, and comparisons under fixed-parameter settings.
For details on other parameters, see \appendixref{app4}.

\begin{table}[!b]
\caption{\textbf{\texttt{AutoClip-DPZO}} utility across estimation schedules on RoBERTa-Large ($\varepsilon=6,\ K=16$).}
\centering
\resizebox{0.95\linewidth}{!}{
\begin{tabular}{c|c||cc}
\hline
\multirow{1}{*}{\textbf{Estimation schedule}}  & \multirow{1}{*}{\textbf{Estimation count}}
& \multicolumn{1}{c}{\textbf{SST-5}}  & \multicolumn{1}{c}{\textbf{MNLI}}  \\
\hline \hline
\multirow{1}{*}{Initialization~(Ours)}   &1    &\cellcolor{gray!40}50.4 &\cellcolor{gray!40}75.2  \\ \hline
\multirow{2}{*}{Uniform}  &2    &50.4 &74.8  \\\cline{2-4}
 &3    &49.6 &74.1  \\  \hline
\multirow{2}{*}{Top-Priority}   &2    &50.2 &74.8  \\ \cline{2-4}
&3
&49.2 &73.6 \\
\hline
\end{tabular}
}
\label{tab9}
\end{table}

\begin{table}[!b]
\caption{Extended results for~\autoref{tab3}: fixed clipping strategies do not generalize across settings ($K=64$).}
\centering
\small
\resizebox{1\linewidth}{!}{
\begin{tabular}{c|cc||cc|cc}
\hline
\multirow{2}{*}{\textbf{Privacy}} &\multicolumn{2}{c||}{\multirow{2}{*}{
\textbf{
Method
}}}  & \multicolumn{4}{c}{\textbf{RoBERTa-Large}}  \\
\cline{4-7}
& & &\multicolumn{1}{c}{\textbf{MNLI}} &\multicolumn{1}{c}{\textbf{SNLI}} & \textbf{SST-2} &\textbf{SST-5}  \\
\hline \hline
& &\multicolumn{3}{c}{--- $K=64$ ---} \\ \cline{2-7}
\multirow{6}{*}{$\varepsilon=3$}
&\multirow{3}{*}{\textbf{Fixed Value}}&\multicolumn{1}{|c||}{$C=4$}  &70.1   &\cellcolor{gray!15}81.1  &92.7  &46.9   \\
& &\multicolumn{1}{|c||}{$C=16$} &68.2   &80.2  &92.8  &48.9     \\
& &\multicolumn{1}{|c||}{$C=32$}  &65.3   &78.1  &90.4  &47.5   \\ \cline{2-7}
& \multirow{3}{*}{\textbf{Fixed Quantile}}  &\multicolumn{1}{|c||}{$Q=0.25$}  &\cellcolor{gray!15}71.8  &80.8 &\cellcolor{gray!15}93.5  &\cellcolor{gray!15}49.1    \\
& &\multicolumn{1}{|c||}{$Q=0.50$}  &68.7  &78.8  &93.4  &48.0      \\
& &\multicolumn{1}{|c||}{$Q=0.75$}  &65.6  &75.5  &93.0  &47.5   \\  \cline{2-7}
&\multicolumn{2}{c||}{\textbf{\texttt{AutoClip-DPZO}}}  &\multicolumn{1}{c}{\cellcolor{gray!40}72.3}
&\multicolumn{1}{c|}{\cellcolor{gray!40}81.6}&\cellcolor{gray!40}93.6
& \cellcolor{gray!40}49.6\\
\hline
\end{tabular}
}
\label{tab10}
\end{table}

\begin{table}[!b]
\caption{Selected $C$ in the main experiments (seed $=42$).}
\centering
\label{tab11}
\resizebox{1\linewidth}{!}{
\begin{tabular}{c|c||cccccc}
\hline
\textbf{Model} &\textbf{Privacy} & \textbf{SST-2} & \textbf{SST-5} & \textbf{SNLI} & \textbf{MNLI} &\textbf{RTE} &\textbf{TREC} \\
\hline
\hline
\multirow{2}{*}{\textbf{RoBERTa-Large}} &\multirow{1}{*}{$\varepsilon=6$}  &4.00  &44.75  &59.75 &52.00 &44.25 &68.50  \\
&\multirow{1}{*}{$\varepsilon=3$}  &0.25  &27.00  &13.25 &14.25 &12.75 &49.75 \\ \hline \hline
& &\multicolumn{2}{c}{\textbf{SST-2}} &\multicolumn{2}{c}{\textbf{SST-5}} &\multicolumn{2}{c}{\textbf{SQuAD}} \\ \hline
\hline
\multirow{2}{*}{\textbf{OPT-350M}} &\multirow{1}{*}{$\varepsilon=6$}  &\multicolumn{2}{c}{19.25} &\multicolumn{2}{c}{36.25} &\multicolumn{2}{c}{25.25} \\
&\multirow{1}{*}{$\varepsilon=3$}  &\multicolumn{2}{c}{2.75} &\multicolumn{2}{c}{27.25} &\multicolumn{2}{c}{15.00} \\
\hline
\end{tabular}
}
\end{table}

\begin{table}[!b]
\caption{RoBERTa runtimes (extending~\autoref{tab4}).
Separately measured runtimes with predefined $C$ are normalized to 1.}
\centering
\small
\resizebox{\linewidth}{!}{
\begin{tabular}{c|c||cc|cc}
\hline
\multirow{3}{*}{\textbf{Task}}
& \multirow{3}{*}{
\textbf{
\begin{tabular}[c]{@{}c@{}}
Model \\
Size
\end{tabular}
}
} & \multicolumn{4}{c}{\textbf{Hardware Time (hours)}}  \\
& & \multicolumn{2}{c}{$K=16$} & \multicolumn{2}{c}{$K=64$} \\
\cline{3-6}
& & Two-Stage & \textbf{\texttt{AutoClip-DPZO}} & Two-Stage & \textbf{\texttt{AutoClip-DPZO}} \\
\hline \hline
\multirow{2}{*}{\textbf{SST-2}}
& \textbf{125M} & 4.20h~(7.23$\times$) & 0.63h~(1.09$\times$) & 8.63h~(7.71$\times$) & 1.32h~(1.18$\times$) \\
& \textbf{355M} & 9.40h~(7.68$\times$) & 1.35h~(1.13$\times$) & 20.57h~(7.76$\times$) & 3.05h~(1.15$\times$) \\
\hline \hline
\multirow{2}{*}{\textbf{RTE}}
& \textbf{125M} & 7.88h~(7.62$\times$) & 1.20h~(1.16$\times$) & 20.20h~(7.10$\times$) & 3.05h~(1.18$\times$) \\
& \textbf{355M} & 20.07h~(7.72$\times$) & 3.10h~(1.19$\times$) & 51.10h~(7.51$\times$) & 7.20h~(1.16$\times$) \\
\hline
\end{tabular}
}
\label{tab12}
\end{table}

\begin{table*}[!t]
\caption{Utility across model sizes: DistilRoBERTa-Base~(D-RoBERTa), OPT-125M, and OPT-1.3B ($K=16$).}
\centering
\resizebox{1\linewidth}{!}{
\begin{tabular}{c|c||cccccc}
\hline
\multirow{2}{*}{\textbf{Privacy constraint}} & \multirow{2}{*}{\textbf{Algorithm}}
& \multicolumn{3}{c}{\textbf{SST-2}}
& \multicolumn{3}{c}{\textbf{SST-5}}  \\
\cline{3-5} \cline{6-8}
& & \textbf{D-RoBERTa}& \textbf{OPT-125M} & \textbf{OPT-1.3B}  & \textbf{D-RoBERTa}& \textbf{OPT-125M} & \textbf{OPT-1.3B} \\
\hline \hline
\multirow{4}{*}{$\varepsilon=3,\ \delta=10^{-5}$}
& Automatic Clipping~\cite{ref18}               &87.7 &82.8  &79.6 &40.7 &57.6 &63.4 \\
& Two-Stage with DPZO  &\cellcolor{gray!15}88.2 &81.5  &87.2 &43.3 &63.6 & 67.0 \\
& Upper Bound~(\autoref{lem2})  &88.0 &\cellcolor{gray!15}83.1  &\cellcolor{gray!15}90.8 &\cellcolor{gray!15}44.0 &\cellcolor{gray!15}64.7 &\cellcolor{gray!15}68.1 \\
& \textbf{\texttt{AutoClip-DPZO}~(Ours)} &\cellcolor{gray!40}88.3 &\cellcolor{gray!40}83.8
&\cellcolor{gray!40}91.2 &\cellcolor{gray!40}44.6
&\cellcolor{gray!40}64.9 &\cellcolor{gray!40}70.9 \\
\hline \hline
\multirow{4}{*}{$\varepsilon=6,\ \delta=10^{-5}$}
& Automatic Clipping~\cite{ref18}               &87.8 &82.9  &87.2 &43.0 &58.5 &67.5 \\
& Two-Stage with DPZO  &\cellcolor{gray!15}88.4 &84.2  & 91.4 &44.8 &\cellcolor{gray!15}65.3 & 71.1 \\
& Upper Bound~(\autoref{lem2})  &87.7 &\cellcolor{gray!15}84.4  &\cellcolor{gray!15}91.6 &\cellcolor{gray!15}45.0 &65.1  &\cellcolor{gray!15}71.3 \\
& \textbf{\texttt{AutoClip-DPZO}~(Ours)} &\cellcolor{gray!40}88.5 &\cellcolor{gray!40}84.8
&\cellcolor{gray!40}92.1 &\cellcolor{gray!40}46.4
&\cellcolor{gray!40}65.9 &\cellcolor{gray!40}73.4 \\
\hline
\end{tabular}
}
\label{tab13}
\end{table*}

\noindent\textbf{Effect of the estimation schedule.}
To investigate the effect of different estimation schedules, we compare initialization, uniform, and top-priority strategies.
Given a total of $T=1000$ training iterations, the initialization strategy (ours) performs estimation only once, at the first iteration.
The uniform strategy evenly distributes the estimation points throughout training: two estimations are conducted at iterations $1$ and $500$, while three estimations are conducted at iterations $1$, $333$, and $666$.
In contrast, the top-priority strategy concentrates the estimation points in the early stage of training, with two estimations at iterations $1$ and $100$, and three estimations at iterations $1$, $100$, and $200$.

Here, \autoref{tab9} demonstrates that none of the above dynamic estimation strategies for $C$ improves model utility compared with our one-time estimation at initialization.
This confirms that such approaches incur additional privacy loss, leaving a smaller privacy budget for the final fine-tuning stage, and that this cost cannot be compensated for by tracking $C$ more frequently.

\noindent\textbf{Extended fixed-strategy results at $K=64$.}
\autoref{tab10} extends the fixed-strategy comparison in \autoref{tab3} to RoBERTa-Large with $K=64$ and $\varepsilon=3$.
\texttt{AutoClip-DPZO} achieves the highest accuracy across all tasks, while fixed strategies generalize poorly.

\noindent\textbf{Actual selected clipping thresholds.}
In addition, for different datasets and settings,~\autoref{tab11} reports the selected clipping thresholds used in the experiments in \autoref{tab2}.
Indeed, since quantile estimation can be performed at a much finer granularity, the selected $C$ often lies between the grid values used by search-based methods.
Furthermore, we perform an additional experiment on TREC with $\varepsilon=3$ using RoBERTa-Large, showing that increasing the clipping threshold from $C=49.75$ to $C=99.50$ reduces the model accuracy from $89.00\%$ to $71.40\%$.
This result confirms our argument regarding the sensitivity of $C$ in \autoref{sec9}.

\begin{table}[htbp]
\caption{Extended results for~\autoref{tab6}: RoBERTa-Base utility across $K$ ($\varepsilon=6$).}
\centering
\resizebox{0.95\linewidth}{!}{
\begin{tabular}{c|c|c||cc}
\hline
\multirow{1}{*}{\textbf{Model}} & \multirow{1}{*}{\textbf{$K$}}  & \multirow{1}{*}{\textbf{Algorithm}}
& \multicolumn{1}{c}{\textbf{SST-2}}  & \multicolumn{1}{c}{\textbf{SST-5}}  \\
\hline \hline
\multirow{6}{*}{\textbf{RoBERTa-Base}}
&\multirow{2}{*}{$32$}  &Upper Bound    &87.2 &35.3  \\
&
& \textbf{\texttt{AutoClip-DPZO}}
&\cellcolor{gray!40}89.3 &\cellcolor{gray!40}41.6 \\ \cline{2-5}
&\multirow{2}{*}{$64$}   &Upper Bound    &89.5 &38.2  \\
&
& \textbf{\texttt{AutoClip-DPZO}}
&\cellcolor{gray!40}89.7 &\cellcolor{gray!40}45.9 \\ \cline{2-5}
&\multirow{2}{*}{$128$} &Upper Bound  &89.0 &43.6   \\
& & \textbf{\texttt{AutoClip-DPZO}}
&\cellcolor{gray!40}89.7 &\cellcolor{gray!40}46.5 \\
\hline
\end{tabular}
}
\label{tab14}
\end{table}

\begin{table}[ht]
\caption{Extended utility results for~\autoref{tab8}: $r_{\mathrm{ref}}$ calibrations ($K=16,\ \varepsilon=6$). \textsuperscript{*} denotes results above the best baseline in~\autoref{tab2}.}
\centering
\setlength{\tabcolsep}{12pt}
\resizebox{\linewidth}{!}{
\begin{tabular}{c|c||ccc}
\hline
\multirow{2}{*}{\textbf{Model}}
& \multirow{2}{*}{\textbf{Dataset}}
& \multicolumn{3}{c}{$r_{\mathrm{ref}}$} \\
\cline{3-5}
&
& median
& GM
& mean \\
\hline
\hline
\multirow{2}{*}{\textbf{RoBERTa-Large}}
& \textbf{SST-2} & $93.7^*$ & $93.8^*$ & $93.8^*$ \\
& \textbf{SST-5} & $50.3^*$ & $50.4^*$ & $50.1^*$ \\
\cline{1-5}
\multirow{2}{*}{\textbf{OPT-350M}}
& \textbf{SST-2} & $88.7^*$ & $88.5^*$ & $88.8^*$ \\
& \textbf{SST-5} & $68.7^*$ & $68.7^*$ & $68.3^*$ \\
\hline
\end{tabular}
}
\label{tab15}
\end{table}

\noindent\textbf{Extended runtime results.}
\autoref{tab12} extends \autoref{tab4} by comparing the runtime of different methods for the RoBERTa model family on SST-2 and RTE under $K=16$ and $K=64$.
Here, we define the unit time as the runtime required to fine-tune the model with a predefined $C$ under the corresponding value of $K$.
\texttt{AutoClip-DPZO} consistently incurs only a trivial overhead of about $9\%$--$19\%$, while the search-based baseline incurs a much larger cost of more than $7\times$ due to repeated fine-tuning.
This gives \texttt{AutoClip-DPZO} a $7.1\times$ speedup over the search-based approach.

\noindent\textbf{Extended results across model sizes.}
\autoref{tab13} presents the impact of model size on test performance across DistilRoBERTa-Base, OPT-125M, and OPT-1.3B under different privacy budgets.
\texttt{AutoClip-DPZO} consistently outperforms all baselines across settings, with utility gains of up to $2.8$ points, further supporting its strong generalizability to models unseen during calibration.
It is worth noting that the Automatic Clipping baseline performs poorly under these settings, further showing that DPZO operates in a different noise regime from DPSGD.
Thus, directly transferring existing methods from DPSGD does not solve the clipping-threshold selection problem in DPZO.
As expected, we observe that larger models tend to achieve better utility within each model family.

\noindent\textbf{Extended results for varying $K$.}
Here, we report additional results for varying $K$ under RoBERTa-Base, which differs from the model used in \autoref{sec20}.
As shown in \autoref{tab14}, our method remains effective for larger values of $K$ compared with the baseline method.

\noindent\textbf{Extended results on $r_{\mathrm{ref}}$ calibration.}
Here, \autoref{tab15} further reports results obtained by calibrating $r_{\mathrm{ref}}$ using the median $(0.257)$ and arithmetic mean $(0.261)$, compared with the geometric mean (GM) $(0.258)$.
As expected, due to the concentration shown in \autoref{fig2}, these calibrations yield similar calibrated values and thus similar model utility.
All three calibration choices outperform the baselines in \autoref{tab2}, demonstrating robustness to the choice of calibration.

\section{Self-Consistent Ladder Calibration}
\label{app3}

We formalize self-consistent calibration and establish its properties to complement \autoref{sec14}.

\subsection{Ladderization and Self-Consistency}
We begin by introducing the ladderization procedure.
Recall that the exact form of $\rho(C)$ is unknown.
Thus, the key idea is to use ladderization to locally freeze $\rho(C)$ within small intervals.
Let $C_q$ denote the clipping threshold corresponding to the $q$-quantile of the gradient norms in a batch, where $q \in [0,1]$.
For $L \geq 2$ ladders of equal width $\delta_q := 1/L$, we define the $\ell$-th ladder as the quantile interval or, equivalently, the clipping-threshold interval:
\begin{equation}
    \label{eq8}
    Q_\ell = ((\ell-1)\delta_q, \ell\delta_q],
    \quad
    I_\ell = (C_{(\ell-1)\delta_q}, C_{\ell\delta_q}].
\end{equation}
This yields a finite set of candidate correction factors $\{\rho_\ell\}_{\ell=1}^L$ to approximate the original $\rho(C)$.
For subsequent analysis, we also define the midpoint quantile level of the $\ell$-th ladder as $q_\ell^{\mathrm{mid}} := (\ell - \tfrac{1}{2})\delta_q$.

The above ladderization procedure yields the ladderized objective $\mathcal{E}_{\mathcal{G}}^{\mathrm{ladder}}(C)$ by replacing $\rho(C)$ with the fixed value $\rho_\ell$ within each ladder:
\begin{equation}
    \label{eq9}
    \mathcal{E}_{\mathcal{G}}^{\mathrm{ladder}}(C)
    = \rho_\ell \mathcal{F}(C) + \lambda C.
\end{equation}
Intuitively, $\mathcal{E}_{\mathcal{G}}^{\mathrm{ladder}}(C)$ approximates $\mathcal{E}_{\mathcal{G}}(C)$, with their difference given by $(\rho_\ell - \rho(C))\mathcal{F}(C)$ within each ladder.
This approximation error can be reduced by increasing the number of ladders.
Therefore, our analysis of $\mathcal{E}_{\mathcal{G}}^{\mathrm{ladder}}(C)$ also applies to $\mathcal{E}_{\mathcal{G}}(C)$, up to an additional approximation error.
In practice, we use 64 ladders to keep the error small.

We now define a self-consistent ladder, whose correction factor is used for our final calibration.
We will later show why this choice yields the desired properties.
Intuitively, self-consistency requires that, when the calibration factor is set to $r = \rho_\ell$ within the $\ell$-th ladder, the proposed quantile that minimizes $\mathcal{E}_{\mathcal{R}}(C)$, $q_\ell^{\mathrm{prop}} = [1-\lambda/\rho_\ell]_{[0,1]}$, also lies within the same quantile interval, i.e., $q_\ell^{\mathrm{prop}} \in Q_\ell$.
We formalize this notion below.

\begin{definition}[Self-consistent ladder]
\label{def1}
    For $0<\lambda<\rho_1(1-q_1^{\mathrm{mid}})$, a ladder $I_\ell$, $\ell\in\{1,\ldots,L\}$, is self-consistent if
    \begin{equation}
\label{eq10}
        \rho_j(1-q_j^{\mathrm{mid}})>\lambda
        \quad\text{for every }j\in\{1,\ldots,\ell\},
    \end{equation}
    and either $\ell=L$ or $\rho_{\ell+1}(1-q_{\ell+1}^{\mathrm{mid}})\le\lambda$.
\end{definition}

Note that the above formal definition slightly extends self-consistency to allow the selected ladder to be only roughly self-consistent due to discretization.
More specifically, consider the case where all ladders on the left propose quantiles to their right, while all ladders on the right propose quantiles to their left.
In this case, the definition selects the rightmost ladder among those on the left as the self-consistent ladder.
Let $\rho_{\mathrm{sc},\phi}$ denote the correction factor of the self-consistent ladder under setting $\phi$.
The resulting calibrated error objective is then given by
\begin{equation*}
    \mathcal{E}_{\mathcal{R}}^{\phi}(C)
    = \rho_{\mathrm{sc},\phi} \cdot \mathcal{F}(C) + \lambda_\phi C.
\end{equation*}

\noindent\textbf{Remark (boundary cases).}
Intuitively, a large noise weight $\lambda$ makes increasing $C$ more costly, pushing the preferred threshold toward the first ladder.
When $\lambda\ge\rho_1(1-q_1^{\mathrm{mid}})$, even the first ladder proposes a small quantile, so we use its coefficient $r=\rho_1$.
With an even larger noise weight, $\lambda\ge\rho_1$, the target reaches the zero quantile.

\subsection{Conservative Threshold Selection}
We show that self-consistent ladder calibration satisfies the desired properties: it gives a tighter threshold than the upper bound while avoiding an overly small threshold.

We first state the monotonicity condition, explain why it naturally holds, and discuss its role in our analysis.

\begin{equation}
    \label{eq11}
    0 < \rho_1 \le \cdots \le \rho_L \le 1.
\end{equation}
Note that, for simplicity, we impose a slightly stronger monotonicity condition in \autoref{sec14}, requiring $\rho(C)$ to be nondecreasing, which implies the condition above.
This assumption is natural: as $C$ increases, stronger, better-aligned gradients are no longer clipped, improving the overall alignment and increasing $\rho(C)$.
Empirical evidence in \appendixref{app5} also supports this assumption.

We now explain why this condition is useful.
The intuition behind this conservative ordering is that, when the alignment factor is nondecreasing, the zeroth-order approximation omits the nonnegative term $\rho'(C)\mathcal{F}(C)$ from the derivative of the alignment-aware objective.
At a critical point of the proxy error objective, its derivative is zero.
Adding back the omitted term therefore makes the alignment-aware derivative nonnegative, suggesting that its optimum lies at or to the left of the proxy threshold.
The following result formalizes this conservative ordering for the ladderized objective.

\begin{theorem}[Conservative self-consistent calibration]
\label{thm2}
    Under \eqref{eq11}, let $\ell$ be the self-consistent ladder defined in \autoref{def1}, and set $r = \rho_\ell$ and $C_{\mathcal{R}}^* = C_{q_\ell^{\mathrm{prop}}}$.
    Then, for the ladderized objective, there exists a threshold $C_{\mathcal{G}}^*$ that is optimal up to a vanishing ladderization error in objective value and satisfies
    \begin{equation}
        \label{eq12}
        C_{\mathcal{G}}^* \le C_{\mathcal{R}}^* \le C_{\mathcal{F}}^*.
    \end{equation}
\end{theorem}

\subsection{Dependence on the Noise Weight \texorpdfstring{$\lambda$}{lambda}}
Finally, we explain why calibration should be performed in the least noisy regime (i.e., with smaller $\lambda$).
The intuition is that a larger $\lambda$ favors stronger clipping, leading to a smaller selected ladder and correction factor.
Calibrating in the least noisy regime therefore yields a larger correction factor, preventing the final calibrated proxy from selecting an overly small threshold.
\autoref{cor1} formalizes this.

\begin{corollary}[Monotonicity in the noise weight]
\label{cor1}
    Fix the batch and the coefficients in \eqref{eq11}.
    For $0<\lambda_1<\lambda_2<\rho_1(1-q_1^{\mathrm{mid}})$, let $i$ and $k$ be the indices of the self-consistent ladders in \autoref{def1}, respectively.
    Then $k\le i$ and $\rho_k\le\rho_i$.
\end{corollary}

The smallest noise weight $\lambda_{\phi_{\min}}$ gives the largest self-consistent correction factor $\rho_{\mathrm{sc},\phi_{\min}}$.
Reusing it at a larger $\lambda_\phi$ gives a threshold no smaller than recalibrating with $\rho_{\mathrm{sc},\phi}$.
This supports using $\rho_{\mathrm{sc},\phi_{\min}}$ as the reference correction factor in \autoref{sec16}.

\section{Additional Experimental Setup}
\label{app4}
In this paper, our primary focus is on optimizing the selection of the DP-specific clipping threshold.
For other learning hyperparameters, we fix the marginal sampling rate to $0.0625$ and use $1,024$ training samples, corresponding to $b=64$.
The procedure for constructing the data subset follows prior work~\cite{ref17}, and we use identical random seeds for sampling the training and test data.
For the learning rate, we specify the search range in \autoref{tab16}.

\begin{table}[!htbp]
\caption{\texttt{AutoClip-DPZO} learning-rate ranges (step $=1.5\times10^{-4}$).}
\centering
\resizebox{0.8\linewidth}{!}{
\label{tab16}
\begin{tabular}{cc||c}
\toprule
\textbf{Dataset} &\textbf{Model Family} & \textbf{Range} \\
\hline \hline
SST-2 &\multirow{6}{*}{RoBERTa} & $[3.5\times10^{-4},\,8\times10^{-4}]$ \\

SST-5 & &$[3.5\times10^{-4},\,8\times10^{-4}]$  \\

SNLI & &$[3.5\times10^{-4},\,1.5\times10^{-3}]$ \\

MNLI & &$[3.5\times10^{-4},\,1.5\times10^{-3}]$  \\

TREC & &$[8\times10^{-4},\,1.5\times10^{-3}]$ \\

RTE & &$[8\times10^{-4},\,1.5\times10^{-3}]$  \\ \hline \hline
SST-2 &\multirow{3}{*}{OPT} &$[5\times10^{-4},\,8\times10^{-4}]$ \\

SST-5 & &$[5\times10^{-4},\,8\times10^{-4}]$  \\
SQuAD & &$[3.5\times10^{-4},\,8\times10^{-4}]$  \\
\bottomrule
\end{tabular}
}
\end{table}

For baseline (vi), described in \autoref{sec20} and \appendixref{app6}, which replaces the fixed $C$ grid with a set of candidate quantiles, we specify for clarity that the quantile grid is $\{0,0.25,0.5,0.75,1\}$.

\section{Additional Empirical Results on the Monotonicity Conditions}
\label{app5}

\begin{figure*}[htbp]
    \centering
    \setlength{\tabcolsep}{2pt}
    \renewcommand{\arraystretch}{1.0}

    \begin{tabular}{ccc}
        \subfloat[RoBERTa-Large, SST-2]{
            \includegraphics[width=0.25\textwidth]{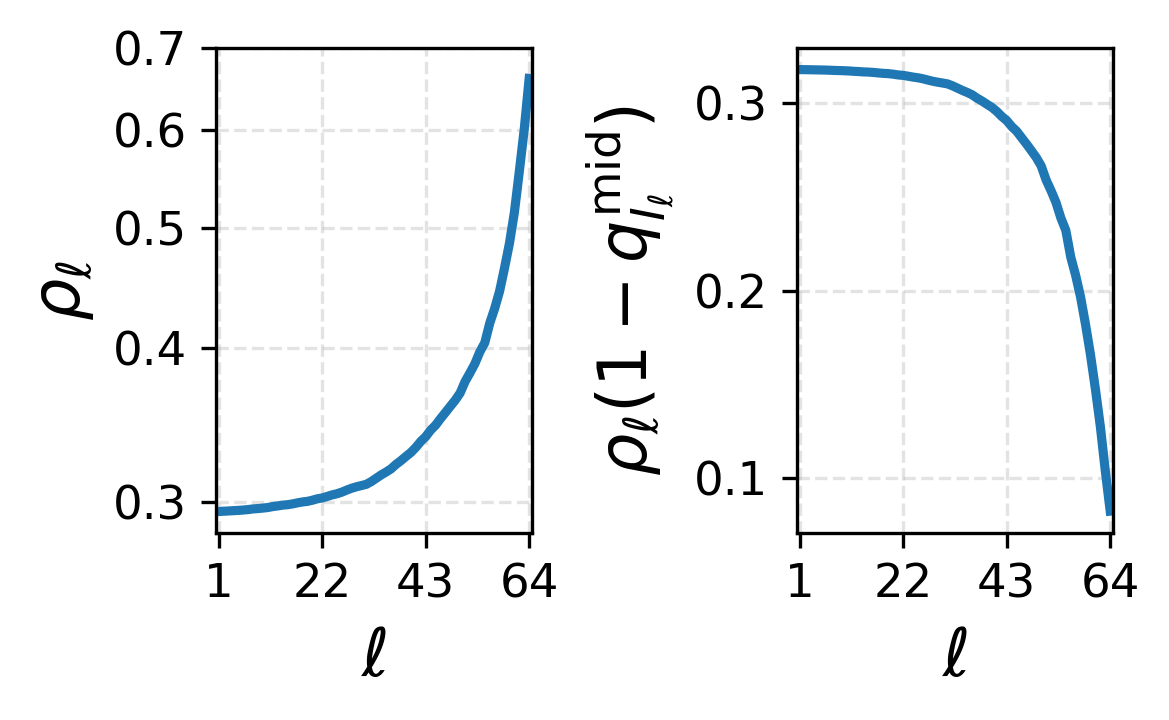}
        }
        &
        \subfloat[RoBERTa-Large, SST-5]{
            \includegraphics[width=0.25\textwidth]{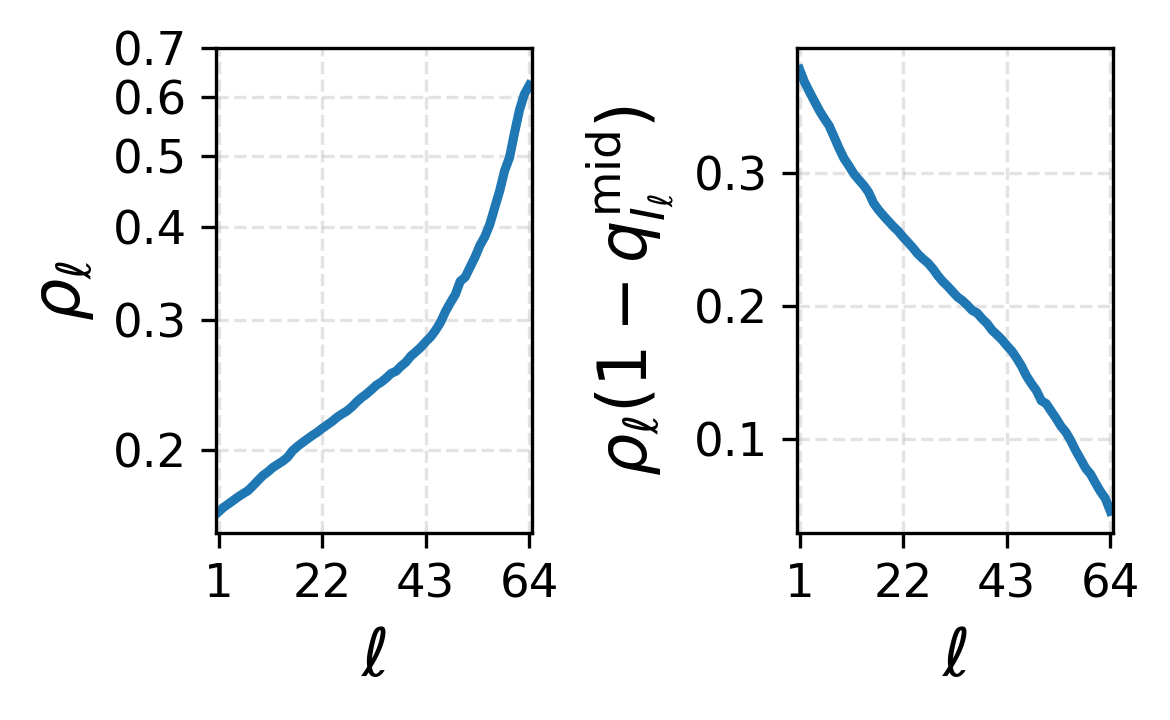}
        }
        &
        \subfloat[RoBERTa-Large, SNLI]{
            \includegraphics[width=0.25\textwidth]{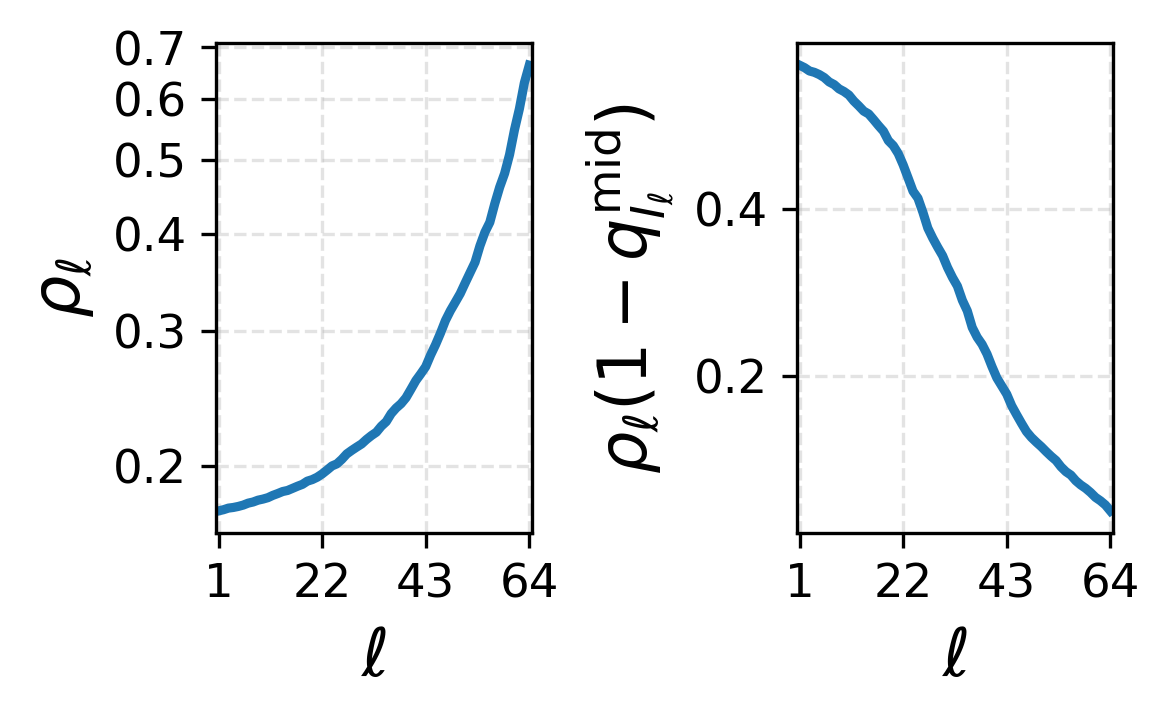}
        }
        \\[0.6em]

        \subfloat[RoBERTa-Large, MNLI]{
            \includegraphics[width=0.25\textwidth]{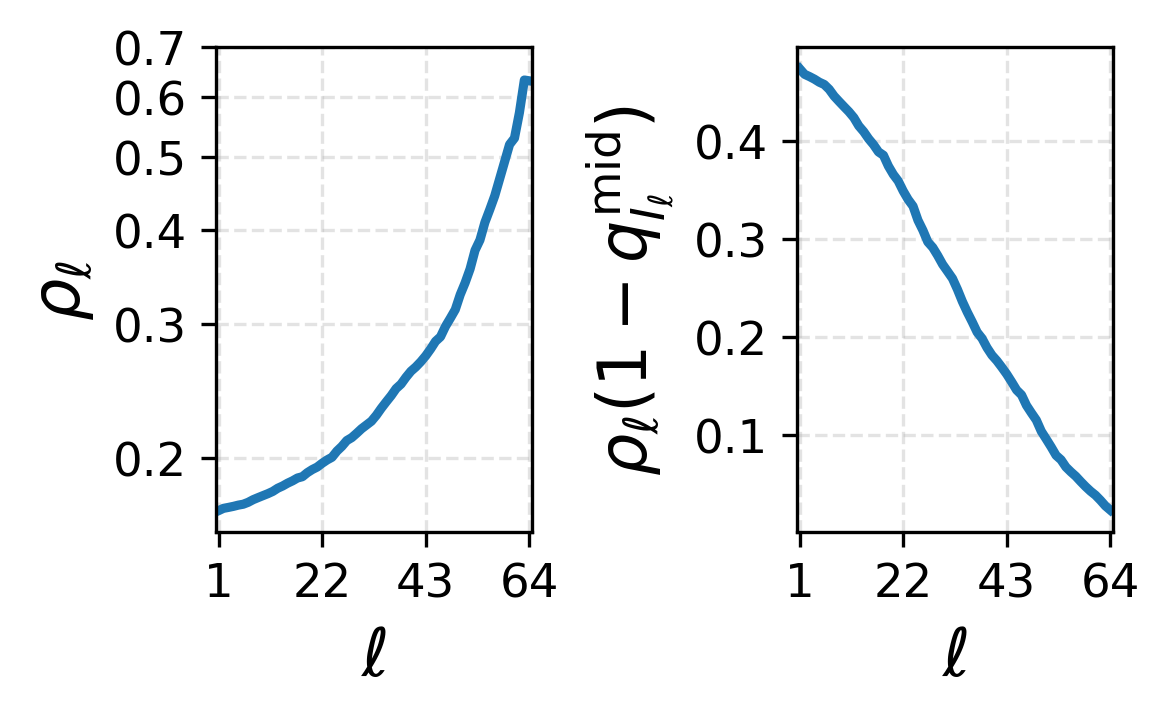}
        }
        &
        \subfloat[RoBERTa-Large, RTE]{
            \includegraphics[width=0.25\textwidth]{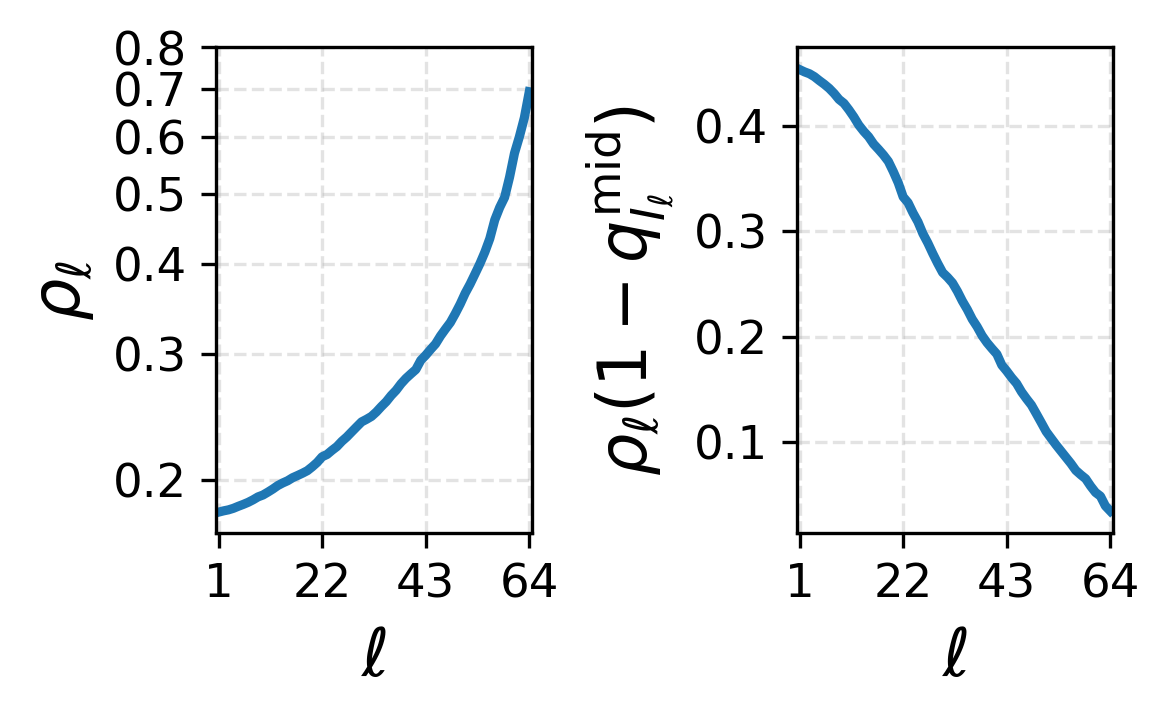}
        }
        &
        \subfloat[RoBERTa-Large, TREC]{
            \includegraphics[width=0.25\textwidth]{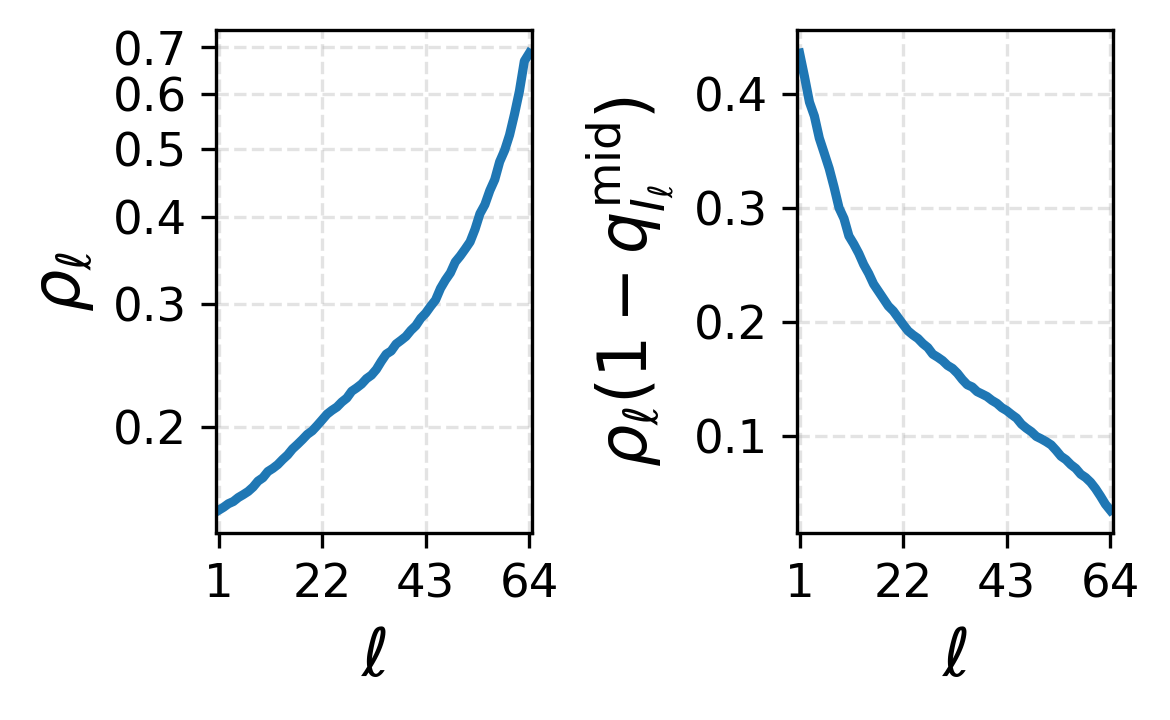}
        }
        \\[0.6em]

        \subfloat[OPT-350M, SST-2]{
            \includegraphics[width=0.25\textwidth]{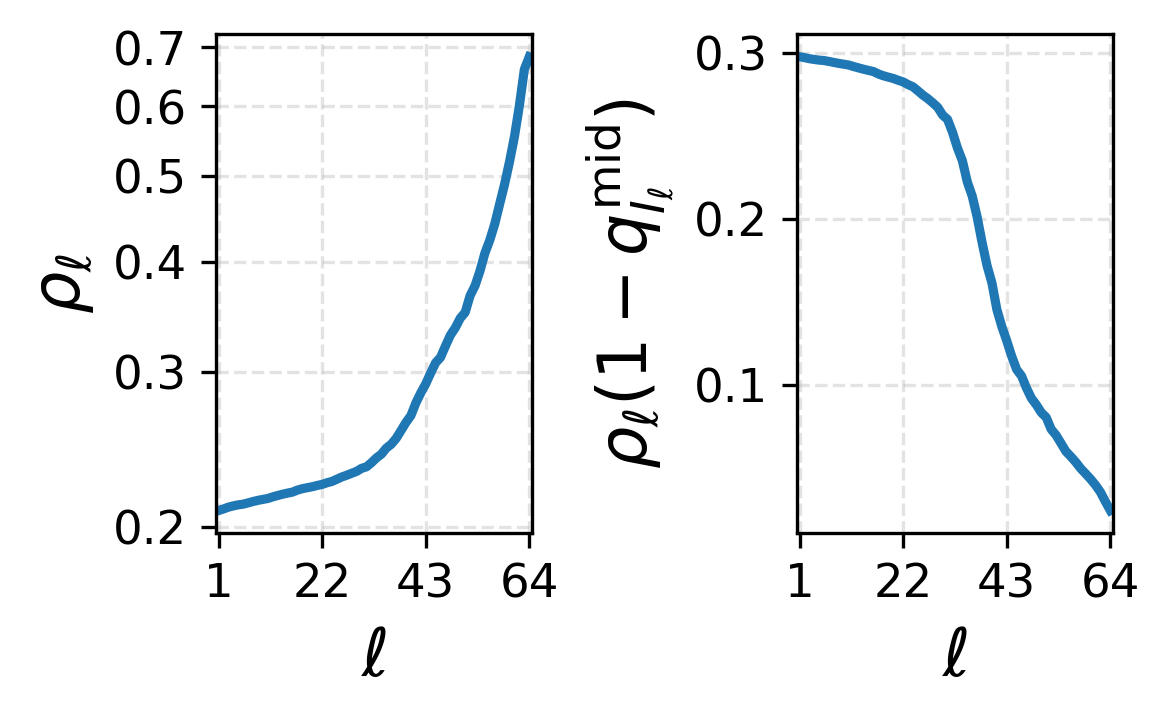}
        }
        &
        \subfloat[OPT-350M, SST-5]{
            \includegraphics[width=0.25\textwidth]{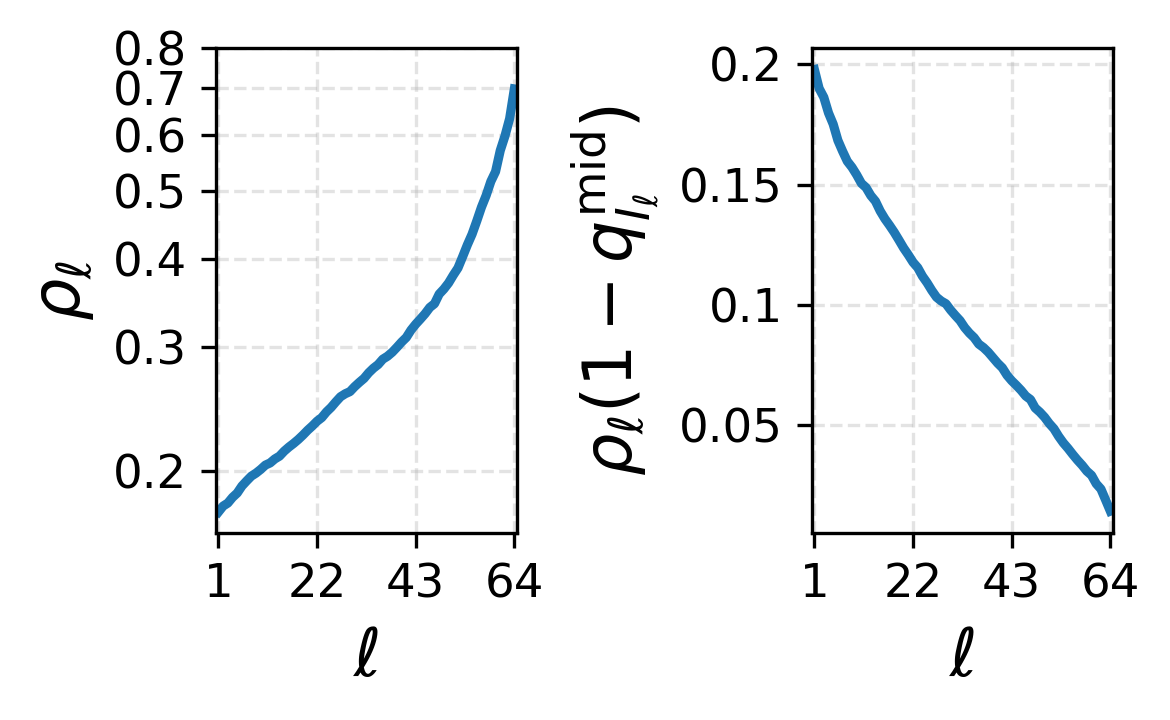}
        }
        &
        \subfloat[OPT-350M, SQuAD]{
            \includegraphics[width=0.25\textwidth]{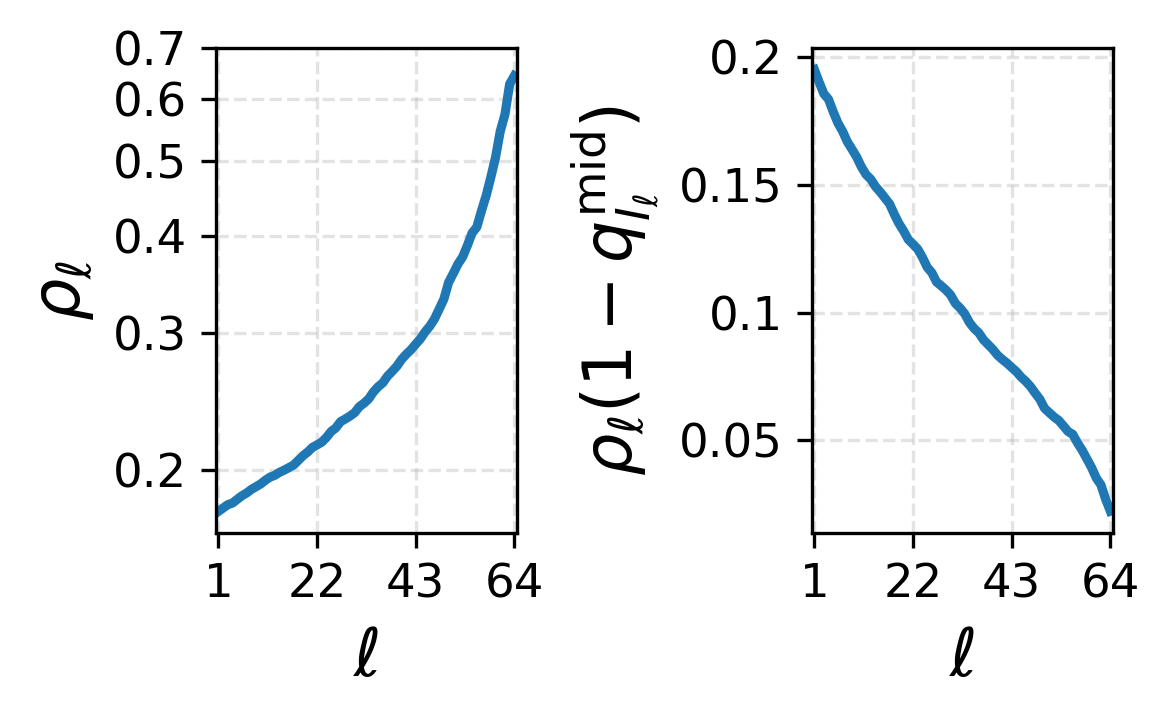}
        }
    \end{tabular}

    \caption{Additional empirical results supporting the monotonicity conditions.}
    \label{fig3}
\end{figure*}

\autoref{fig3} shows the ladderized $\rho_\ell$ values under a single random seed, illustrating the monotonicity required by \autoref{thm1} and \autoref{thm2}.
In contrast, the final calibration of $r_{\mathrm{ref}}$ uses batches sampled with three random seeds.
Here, we use $64$ ladders and perform the experiment at the initial model checkpoint, averaging over $64$ random batches with $K=16$ and $b=64$.
It demonstrates that these functions have similar shapes, which helps explain the concentration of the calibration factor observed in \autoref{sec16}.
This experiment further supports, from an empirical perspective, the intuitive justification developed in the high-level discussion in \autoref{sec14} and \appendixref{app3}.
Please note that, because both data points and random directions are drawn i.i.d., these observations remain stable in expectation as long as the batch size and the number of directions are not too small.

\section{The Formal Algorithm of the Two-Stage Strawman Protocol}
\label{app6}
\autoref{alg2} formalizes the idea in \autoref{sec10}.
Here, for simplicity, $\mathcal{M}$ denotes the DPZO algorithm, and $\mathrm{Evaluate}(\cdot)$ denotes a possibly task-dependent model-utility criterion for comparing candidate models.

\begin{algorithm}[htbp]
\caption{Two-Stage Strawman Protocol}
\label{alg2}
\LinesNumbered
\SetAlgoLined
\DontPrintSemicolon
\SetNoFillComment
\KwIn{Dataset $D$, initial model $\theta_{\mathrm{init}}$, number of directions $K$, batch size $b$, number of \emph{search} steps per candidate $T_1$, number of \emph{release} steps $T_2$, total privacy budget $(\varepsilon,\delta)$, allocation factor $\alpha \in (0,1)$, candidate set $\{C_m\}_{m=1}^M$}
\KwOut{Final privately fine-tuned model parameters $\theta^*$}

\tcp{Stage 1: The search stage}
$\varepsilon_{\mathrm{search}} \gets \alpha \varepsilon,\quad
\delta_{\mathrm{search}} \gets \alpha \delta$\;
$u_{\mathrm{best}} \gets -\infty,\quad
C^* \gets C_1$\;

$\sigma_{\mathrm{search}} \gets
\mathsf{MomentAccount}(\varepsilon_{\mathrm{search}}, \delta_{\mathrm{search}}, \frac{b}{|D|}, T_1 M)$\;

\For{$m = 1, \dots, M$}{
    $\theta_m \gets \mathcal{M}(D, \theta_{\mathrm{init}}, K, b, T_1, C_m, \sigma_{\mathrm{search}})$\;
    \tcp*{Model evaluation on the test set}
    $u_m \gets \mathrm{Evaluate}(\theta_m)$\;

    \If{$u_m > u_{\mathrm{best}}$}{
        $u_{\mathrm{best}} \gets u_m,\quad C^* \gets C_m$\;
    }
}

\tcp{Stage 2: The release stage}
$\varepsilon_{\mathrm{release}} \gets (1-\alpha)\varepsilon,\quad
\delta_{\mathrm{release}} \gets (1-\alpha)\delta$\;

$\sigma_{\mathrm{release}} \gets
\mathsf{MomentAccount}(\varepsilon_{\mathrm{release}}, \delta_{\mathrm{release}}, b/|D|, T_2)$\;

$\theta^* \gets \mathcal{M}(D, \theta_{\mathrm{init}}, K, b, T_2, C^*, \sigma_{\mathrm{release}})$\;

\Return $\theta^*$\;
\end{algorithm}

In \autoref{sec20}, we further discuss baseline (vi), which replaces the candidate set $\{C_m\}_{m=1}^M$ in the above algorithm with a set of quantiles $\{q_m\}_{m=1}^M$.
One may then derive a similar method that first constructs $D_{\text{est}}$ as in \autoref{alg1} and then uses private quantile estimation to obtain a data-dependent candidate set $\{C_m\}_{m=1}^M$.
This enables a more ``data-aware'' approach and can be viewed as an improved version of~\cite{ref43}, which considers only the median, i.e., the $50\%$ quantile. However, this approach incurs the additional cost of sequentially composing all private quantile estimation procedures.

\section{Proofs of Lemmas, Theorems, and Corollaries}
\label{app7}

Here, we present the proofs of the new lemmas, theorems, and corollaries introduced in this work.

{
\subsection{Proof of \autoref{lem2}}
\label{proof1}

\begin{proof}
    From \eqref{eq4}, it suffices to minimize
    \begin{equation*}
    \mathcal{E}_\mathcal{F}(C)=\frac{1}{b}\left(\sum_{x \in B} \max\!\left(0, \left\| g_x \right\|_2 - C\right) + \sigma C \sqrt{K}\right).
    \end{equation*}
    Away from the sample norms, differentiating with respect to $C$ gives
    \begin{equation*}
        \frac{\partial}{\partial C}\mathcal{E}_\mathcal{F}(C)
        = -\frac{1}{b} \sum_{x \in B} \mathbb{I}\!\left(\|g_x\|_2 > C\right) + \frac{\sigma \sqrt{K}}{b}.
    \end{equation*}
    Therefore, a positive differentiable minimizer $C_\mathcal{F}^*$ satisfies
    \begin{equation*}
        \sum_{x \in B} \mathbb{I}\!\left(\|g_x\|_2 > C_\mathcal{F}^*\right) = \sigma \sqrt{K}.
    \end{equation*}
    Equivalently, the fraction of samples whose norm is at most $C_\mathcal{F}^*$ is $1 - \frac{\sigma \sqrt{K}}{b}$. Hence,
    \begin{equation*}
        C_\mathcal{F}^* = \operatorname{Quantile}\!\left( \{\|g_x\|_2\}_{x \in B}, \, \left[1-\frac{\sigma\sqrt{K}}{b}\right]_{[0,1]} \right).
    \end{equation*}
    \emph{Remark.}
    It is straightforward to see that when $\sigma\sqrt{K}$ is not an integer, the derivative cannot be zero.
    Nevertheless, the target error is minimized by selecting the smallest positive threshold such that at least the target fraction of sample norms lie at or below it.
    Thus, the quantile rule still applies.
    Furthermore, we exclude $C=0$, as it would eliminate the gradient signal entirely.
\end{proof}
\subsection{Proof of \autoref{lem3}}
\label{proof2}
\begin{proof}
    From \eqref{eq7}, since $r>0$, it suffices to minimize
    \begin{equation*}
        \frac{\mathcal{E}_{\mathcal{R}}(C)}{r}
        = \frac{1}{b}\sum_{x\in B}\max\!\left(0,\|g_x\|_2-C\right)
        + \frac{\lambda}{r}C.
    \end{equation*}
    This has the same form as the upper-bound objective, with $\lambda$ replaced by $\lambda/r$. Hence, by \autoref{lem2},
    \begin{equation}
    \label{eq13}
        C_{\mathcal{R}}^*
        = \operatorname{Quantile}\!\left(\{\|g_x\|_2\}_{x\in B},\,\left[1-\frac{\lambda}{r}\right]_{[0,1]}\right).
    \end{equation}
    Since $0<r\le1$, the induced quantile level is no larger than $[1-\lambda]_{[0,1]}$, so $C_{\mathcal{R}}^*\le C_{\mathcal{F}}^*$.
    Note that the remark on the non-integer case in~\appendixref{proof1} also applies here.
\end{proof}
\subsection{Proof of \autoref{thm2}}
\begin{proof}
    Although the following proofs are somewhat technical, we first provide the intuition behind them.
    Since $C_{\mathcal{R}}^* \le C_{\mathcal{F}}^*$ follows directly for $r \in (0,1]$, we focus on proving the first inequality, i.e., $C_{\mathcal{G}}^* \le C_{\mathcal{R}}^*$.
    Under the extended definition of self-consistency, the selected ladder may propose a quantile in the next ladder, so we first locate the corresponding threshold $C_{\mathcal{R}}^*$ explicitly.
    We then show that any improvement to the right of $C_{\mathcal{R}}^*$ is bounded by a ladderization error that vanishes as the number of ladders increases.
    Therefore, there exists a threshold $C_{\mathcal{G}}^* \le C_{\mathcal{R}}^*$ whose objective value is optimal up to the vanishing ladderization error, showing the desired ordering.

    \emph{Step 1: locate $C_{\mathcal{R}}^*$.}
    Let $I_\ell$ be the ladder selected for calibration, and let $I_j$ be the ladder containing $C_{\mathcal{R}}^*$.
    \begin{enumerate}
        \item If $\ell<L$, we show that $C_{\mathcal{R}}^*\ge C_{(\ell-1)\delta_q}$ and $j\le\ell+1$.
        \item If $\ell=L$, we show that $C_{\mathcal{R}}^*\ge C_{(L-1)\delta_q}$ and $j\le L$.
    \end{enumerate}

    \emph{First, suppose the calibrated ladder has index $\ell<L$.}
    By \autoref{def1}, $\rho_\ell(1-q_\ell^{\mathrm{mid}})>\lambda$.
    Dividing by $\rho_\ell>0$ and rearranging gives $q_\ell^{\mathrm{mid}}<1-\lambda/\rho_\ell=q_\ell^{\mathrm{prop}}$.
    The same argument applies to ladder $\ell+1$. Here the definition gives $\rho_{\ell+1}(1-q_{\ell+1}^{\mathrm{mid}})\le\lambda$, so $1-\lambda/\rho_{\ell+1}\le q_{\ell+1}^{\mathrm{mid}}$.
    Combining the two midpoint bounds with the coefficient monotonicity condition $\rho_\ell\le\rho_{\ell+1}$ from~\eqref{eq11} gives
    \begin{equation}
    \label{eq14}
        \begin{aligned}
            q_\ell^{\mathrm{mid}} < q_\ell^{\mathrm{prop}}
            &= 1-\frac{\lambda}{\rho_\ell}\\
            &\le 1-\frac{\lambda}{\rho_{\ell+1}}
            \le q_{\ell+1}^{\mathrm{mid}}.
        \end{aligned}
    \end{equation}
    The two midpoints in~\eqref{eq14} lie between the lower edge $(\ell-1)\delta_q$ of ladder $\ell$ and the upper edge $(\ell+1)\delta_q$ of ladder $\ell+1$.
    The proposed quantile thus lies between these edges as well. Since the clipping threshold is nondecreasing in $q$, we have $C_{(\ell-1)\delta_q}\le C_{\mathcal{R}}^*\le C_{(\ell+1)\delta_q}$.
    Because the right-hand side $C_{(\ell+1)\delta_q}$ is the upper boundary of ladder $\ell+1$, the ladder containing $C_{\mathcal{R}}^*$ has index $j\le\ell+1$.
    This shows the results in case~(1).

    \emph{Second, suppose the calibrated ladder has index $\ell=L$.}
    Applying the same midpoint calculation to ladder $L$ gives $q_L^{\mathrm{prop}}>q_L^{\mathrm{mid}}>(L-1)\delta_q$, hence $C_{\mathcal{R}}^*\ge C_{(L-1)\delta_q}$. Since $L$ is the last ladder, $j\le L$.
    This shows the results in case~(2).

    In either case, we have proved that $C_{\mathcal{R}}^*\ge C_{(\ell-1)\delta_q}$.
    Since $C_{(\ell-1)\delta_q}$ is the lower boundary of $I_\ell$, every earlier ladder ends at or below $C_{\mathcal{R}}^*$.
    Intuitively, this result suggests that calibration keeps the proposed quantile within the selected ladder or the next one.

    \emph{Step 2: compare larger thresholds.}
    We now show that selecting a clipping threshold larger than $C_{\mathcal{R}}^*$ cannot further reduce the error objective, up to a ladderization error.

    Consider any $C>C_{\mathcal{R}}^*$.
    If $C$ lies in ladder $I_k$, Step~1 gives $k\ge\ell$, so $\rho_k\ge\rho_\ell$ by~\eqref{eq11}.
    The ladderized objective $\mathcal{E}_{\mathcal{G}}^{\mathrm{ladder}}(C)$ and the proxy $\mathcal{E}_{\mathcal{R}}(C)$ share the noise term $\lambda C$, while their bias terms are $\rho_k\mathcal{F}(C)$ and $\rho_\ell\mathcal{F}(C)$, respectively.
    Since $\mathcal{F}(C)\ge0$, the ladderized objective is no smaller than the proxy.
    Above $C_1$, no clipping occurs, so $\mathcal{F}(C)=0$ and the objectives coincide.
    Finally, the proxy is minimized at $C_{\mathcal{R}}^*$, so
    \begin{equation}
    \label{eq15}
        \begin{aligned}
            \mathcal{E}_{\mathcal{G}}^{\mathrm{ladder}}(C)
            &\ge \mathcal{E}_{\mathcal{R}}(C)
            \ge \mathcal{E}_{\mathcal{R}}(C_{\mathcal{R}}^*)\\
            &= \mathcal{E}_{\mathcal{G}}^{\mathrm{ladder}}(C_{\mathcal{R}}^*)
            +(\rho_\ell-\rho_j)\mathcal{F}(C_{\mathcal{R}}^*).
        \end{aligned}
    \end{equation}
    As shown in the first step, $j\le\ell+1$. If $j\le\ell$, then $\rho_j\le\rho_\ell$ by monotonicity, and \eqref{eq15} shows that no larger threshold improves the objective.
    It therefore remains to bound the possible improvement when $j=\ell+1$, which is at most $(\rho_{\ell+1}-\rho_\ell)\mathcal{F}(C_{\mathcal{R}}^*)$.

    \emph{Step 3: bound the ladderization error.}
    When $j=\ell+1$, we have $q_\ell^{\mathrm{prop}}>\ell\delta_q$, so the proposed quantile is within half a ladder width of the next midpoint.
    The midpoint condition and $\lambda=\rho_\ell(1-q_\ell^{\mathrm{prop}})$ give
    \begin{equation}
    \label{eq16}
        \begin{aligned}
            (\rho_{\ell+1}-\rho_\ell)(1-q_\ell^{\mathrm{prop}})
            &\le \rho_{\ell+1}(q_{\ell+1}^{\mathrm{mid}}-q_\ell^{\mathrm{prop}})\\
            &\le \delta_q/2.
        \end{aligned}
    \end{equation}
    Furthermore, at most a fraction $1-q_\ell^{\mathrm{prop}}$ of sample norms exceed $C_{\mathcal{R}}^*$, each by at most $C_1$.
    Hence, $\mathcal{F}(C_{\mathcal{R}}^*) \le C_1(1-q_\ell^{\mathrm{prop}})$, and the possible improvement in the objective is bounded by $C_1\delta_q/2 = C_1/(2L)$.
    Clearly, this error can be made arbitrarily small by increasing $L$.

    More formally, let $C_{\mathcal{G}}^*$ minimize the objective over $[0,C_{\mathcal{R}}^*]$.
    The preceding bound suggests
    \begin{equation}
    \label{eq17}
        \mathcal{E}_{\mathcal{G}}^{\mathrm{ladder}}(C_{\mathcal{G}}^*)
        \le \min_{C\ge0}\mathcal{E}_{\mathcal{G}}^{\mathrm{ladder}}(C)
        +\frac{C_1}{2L}.
    \end{equation}
    Here, the possible gap arises precisely from the case discussed in the intuition at the beginning of this proof, where the proposed threshold falls in the next ladder.
    This gap vanishes as $L$ increases.
    Additionally, if $q_\ell^{\mathrm{prop}} \in Q_\ell$, then $C_{\mathcal{G}}^*$ is a global minimizer, as follows from~\eqref{eq15}.
\end{proof}
\subsection{Proof of \autoref{cor1}}
\begin{proof}
    The conditions in \autoref{def1} equivalently give
    \begin{equation*}
        \begin{aligned}
            \ell
            &=\min\Bigl(
                \bigl\{j\in\{1,\ldots,L\}:\\
            &\qquad
                \rho_j(1-q_j^{\mathrm{mid}})\le\lambda\bigr\}
                \cup\{L+1\}\Bigr)-1.
        \end{aligned}
    \end{equation*}
    Every index $j$ satisfying $\rho_j(1-q_j^{\mathrm{mid}})\le\lambda_1$ also satisfies the same inequality with $\lambda_2$.
    Including the common element $L+1$, the minimum for $\lambda_2$ is no larger.
    Subtract one, then use coefficient monotonicity.
\end{proof}
}

\section{Additional DP Background}
\label{app8}

DP satisfies sequential composition as follows:

\begin{lemma}[Sequential Composition of Standard DP~\cite{ref9}]
\label{lem4}
    Let $\{\mathcal{M}_1,\, \ldots,\, \mathcal{M}_k\}$ be a series of $(\varepsilon_i,\delta_i)$-DP mechanisms.
    Then $\mathcal{M}_{[k]}$, defined as
    \begin{equation*}
        \mathcal{M}_{[k]}(D) := (\mathcal{M}_1(D),\, \mathcal{M}_2(D),\, \ldots,\, \mathcal{M}_k(D)),
    \end{equation*}
    satisfies $(\varepsilon',\delta')$-DP, where $\varepsilon' = \sum_{i=1}^{k} \varepsilon_i$ and $\delta' = \sum_{i=1}^{k} \delta_i$.
\end{lemma}

For model training, where the Gaussian mechanism is applied iteratively to Poisson-sampled subsets of the data, the Moments Accountant can provide a much tighter composition guarantee than sequential composition.

\begin{theorem}[Composed Subsampled Gaussian Mechanism via the Moments Accountant~\cite{ref11,ref27}]
There exist constants $c_1$ and $c_2$ such that, given the marginal sampling probability $p=b/n$ and the number of steps $T$, for any $\varepsilon < c_1 p^2 T$, the composition of the subsampled Gaussian mechanism over $T$ steps satisfies $(\varepsilon,\delta)$-DP for any $\delta > 0$ if
\begin{equation*}
    \sigma \geq c_2 \frac{b\sqrt{T\log(1/\delta)}}{n\varepsilon}.
\end{equation*}
\end{theorem}

This composition guarantee shows that the noise multiplier in the Gaussian mechanism scales smoothly with the number of training steps.
It was first introduced in the pioneering work on DPSGD~\cite{ref11} under Poisson sampling.
Later work adopted \textit{Rényi differential privacy} (RDP)~\cite{ref63,ref27}, a widely used privacy model for iterative algorithms based on \textit{Rényi divergence}, to provide a more general form of privacy accounting for subsampled mechanisms beyond the Gaussian mechanism and for sampling without replacement, with a similar form.
We do not cover those extensions in this study.

\noindent\textbf{Remark:}
Throughout this work, we use sequential composition between the $C$-selection stage and the final DPZO fine-tuning stage, while consistently applying the standard moments accountant across fine-tuning iterations to achieve a tight privacy accounting bound.

\subsection{Private Quantile Estimation}
\label{app9}

A widely used $(\varepsilon,0)$-DP mechanism is the \textit{exponential mechanism}~\cite{ref23}, which selects an output from a finite domain by sampling according to a data-dependent probability distribution.
Let $\mathcal{Y}$ be a discrete output range and let $\phi: \mathcal{D} \times \mathcal{Y} \to \mathbb{R}$ be a score function with sensitivity
\begin{equation*}
    \Delta(\phi) := \max_{D \sim D'} \max_{y \in \mathcal{Y}} \left\| \phi(D,y) - \phi(D',y) \right\|_1.
\end{equation*}
The exponential mechanism outputs $y \in \mathcal{Y}$ with probability proportional to $\exp\!\left(\frac{\varepsilon \cdot \phi(D,y)}{2\Delta(\phi)}\right)$.
Intuitively, the exponential mechanism privatizes the output using a problem-tailored randomized response rather than noise injection, and thus enjoys the following strong utility guarantees.

\begin{lemma}[Utility of the exponential mechanism~\cite{ref23}]
\label{lem5}
    For any dataset $D$, any $\varepsilon > 0$, and any failure probability $\beta \in (0,1)$, let $\hat{y} \in \arg\max_{y \in \mathcal{Y}} \phi(D,y)$.
    Then, with probability at least $1-\beta$, the exponential mechanism outputs some $y \in \mathcal{Y}$ satisfying
    \begin{equation*}
        \phi(D,y) \ge \phi(D,\hat{y}) - \frac{2\Delta(\phi)}{\varepsilon}\ln\left(\frac{|\mathcal{Y}|}{\beta}\right).
    \end{equation*}
\end{lemma}
It has been shown that the exponential mechanism is particularly effective for selecting the $\tau$-rank quantile under DP~\cite{ref26}, while also enjoying the utility guarantee in \autoref{lem5}.
Informally,~\cite{ref26} achieves a rank error of $O(1/\varepsilon\cdot\log(|\mathcal{Y}|/\beta))$, where $\beta$ denotes the failure probability of the error bound and is typically set to $10\%$.
We denote this mechanism by $\mathsf{Quantile}(D,\mathcal{Y},\varepsilon,\tau)$.
The full procedure is given in \autoref{alg3}.

\begin{algorithm}[htbp]
    \caption{\textsc{FiniteDomainQuantile}~\cite{ref26}}
    \label{alg3}
    \LinesNumbered
    \SetAlgoLined
    \DontPrintSemicolon
    \SetNoFillComment

    \KwIn{$D$, $\mathcal{Y}$, $\varepsilon$, $q$, $\beta=10\%$}
    \KwOut{$\tilde{y} \in \mathcal{Y}$}
    $\tau=q\cdot|D|$

    $\tau'
    =
    \min\!\left\{
        \max\!\left\{
            \tau,\;
            \frac{2}{\varepsilon}\log\!\left(\frac{|\mathcal{Y}|}{\beta}\right)
        \right\},\;
        |D| - \frac{2}{\varepsilon}\log\!\left(\frac{|\mathcal{Y}|}{\beta}\right)
    \right\}$\;

    \ForEach{$y \in \mathcal{Y}$}{
        $w(y) \gets \exp\!\left(
        -\frac{\varepsilon
        \left|
            \tau' - \operatorname{rank}(y)
        \right|}{2}
        \right)$ \tcp*{calculate weights}
    }

    $p(y) \gets \frac{w(y)}{\sum_{y' \in \mathcal{Y}} w(y')}, \quad \forall y \in \mathcal{Y}$ \tcp*{normalize}

    Sample $\tilde{y} \sim p(\cdot)$\;
    \Return $\tilde{y}$\;
\end{algorithm}

{
\makeatletter
\setlength{\@fptop}{0pt}
\makeatother
\clearpage
}

\end{document}